\documentclass[preprint]{elsarticle}
\usepackage{amssymb,amsmath,amsfonts}
\usepackage[numbers]{natbib}
\usepackage{color}
\usepackage[english]{babel}
\usepackage{float}

\usepackage[]{graphicx}

\newcommand{\bfE}{\mathbf{E}}
\newcommand{\bE}{\mathbf{E}}
\newcommand{\bfB}{\mathbf{B}}
\newcommand{\bB}{\mathbf{B}}

\newcommand{\bfv}{\mathbf{v}}
\newcommand{\bfx}{\mathbf{x}}
\newcommand{\bv}{\mathbf{v}}
\newcommand{\bx}{\mathbf{x}}
\newcommand{\bfxi}{{\boldsymbol \xi}}
\newcommand{\bfj}{\mathbf{j}}

\journal{JCP}

\floatstyle{boxed}
\restylefloat{figure}
\begin{document}

\begin{frontmatter}

\title{A Multi Level Multi Domain Method for Particle In Cell Plasma Simulations}

\author[KUL]{M.E. Innocenti\corref{cor1}}
\ead{mariaelena.innocenti@wis.kuleuven.be}
\cortext[cor1]{Corresponding author}

\author[KUL,EXA,other]{G. Lapenta}
\ead{giovanni.lapenta@wis.kuleuven.be}

\author[KUL,EXA]{S. Markidis}
\ead{stefano.markidis@wis.kuleuven.be}

\author[KUL,EXA]{A. Beck}
\ead{arnaud.beck@wis.kuleuven.be}

\author[KUL,EXA]{A. Vapirev}
\ead{alexander.vapirev@wis.kuleuven.be}

\address[KUL]{Center for Plasma Astrophysics, Department of Mathematics, K.U.Leuven, Celestijnenlaan 200B, B-3001 Leuven, Belgium.}
\address[EXA]{ExaScience Intel Lab Europe, Kapeldreef 75, B-3001 Leuven, Belgium}
\address[other]{Space Science, Technology, and Applications (LASA), Leuven Mathematical Modeling \& Computational Science Research Center (LMCC), K.U.Leuven, Celestijnenlaan 200B, B-3001 Leuven, Belgium.}

\begin{abstract}
A novel adaptive technique for electromagnetic Particle In Cell (PIC) plasma simulations is presented here. Two main issues are identified in designing adaptive techniques for PIC simulation: first, the choice of the size of the particle shape function in progressively refined grids, with the need to avoid the exertion of self-forces on particles, and, second, the necessity to comply with the strict stability constraints of the explicit PIC algorithm. The adaptive implementation presented responds to these demands with the introduction of a Multi Level Multi Domain (MLMD) system (where a cloud of self-similar domains is fully simulated with both fields and particles) and the use of an Implicit Moment PIC method as baseline algorithm for the adaptive evolution. Information is exchanged between the levels with the projection of the field information from the refined to the coarser levels and the interpolation of the boundary conditions for the refined levels from the coarser level fields. Particles are bound to their level of origin and are prevented from transitioning to coarser levels, but are repopulated at the refined grid boundaries with a splitting technique. The presented algorithm is tested against a series of simulation challenges.
\end{abstract}


\end{frontmatter}
\section{Introduction}
\label{sec:intro}
\begin{figure}[ht]
\centering
 \includegraphics[width=6cm]{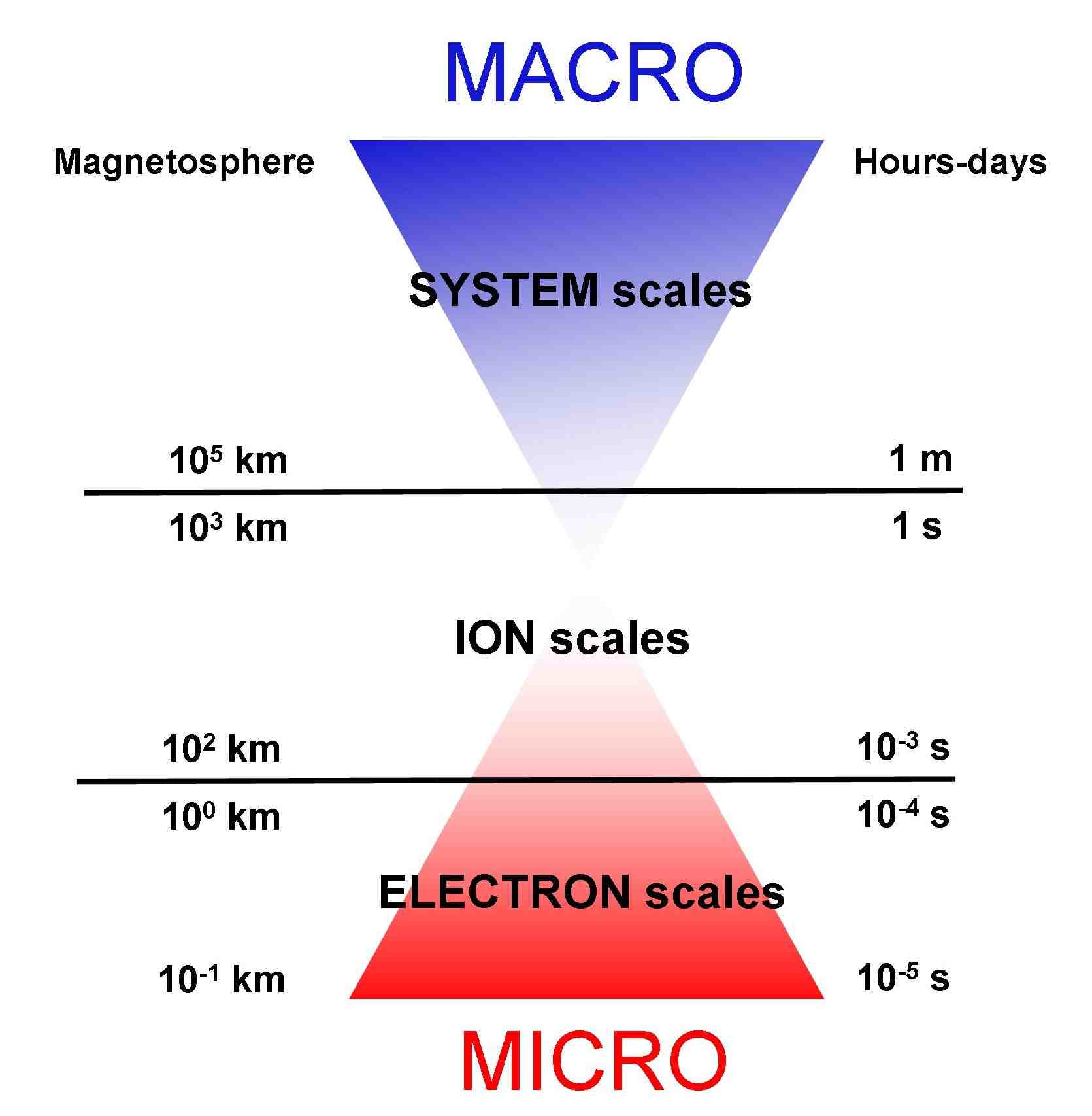}
\caption{Representation of the typical plasma scales observed in the Earth magnetotail. Notice the wide range of spatial and temporal scales which makes the simulation of the relevant phenomena happening in such a system an extremely expensive task under the computational point of view.}
\label{scales}
\end{figure}

The fascinating task of plasma physics simulations is cursed with the coexistence of multiple space and time scales in all relevant phenomena. As an example, Fig.~\ref{scales} reports the typical scales of the plasmas populating different regions of the Earth magnetotail. Notice that the smallest scales are of the order of the tens of microseconds and of the hundreds of meters, considerably smaller than those of interest for the most important processes observed in space, such as the geomagnetic storms caused by the arrival of Coronal Mass Ejections (CME) at Earth or the substorms originated in the Earth environment by magnetic reconnection in the magnetotail region\citep{gonzales_storms}. Those events involve the whole Earth environment and last for minutes to hours to days, making their simulation impossible if all plasma scales are to be encompassed.\\ A common strategy to make these simulations feasible with the computational resources currently available is to neglect the smaller, kinetic scales and use fluid\citep{multiscale} of hybrid (fluid electrons, kinetic ions)\citep{multiscale,dietmar} approaches, at the expense of complete consistency and of the kinetic effects.

If such effects need to be retained, explicit Particle In Cell (PIC) methods~\citep{birdsall} are one of the most frequently employed options. However, their strict stability constraints (Ref.~\cite{birdsall} and~\cite{hockney}) impose notable consequences for space simulations~\citep{lapenta2012_spaceWeather}, requiring for the smallest Debye length (an intrinsic plasma scale measuring the shielding length in a plasma~\citep{hockney,birdsall}) and the fastest electron plasma frequency to be resolved. This task is particularly challenging since often space plasmas present regions of high density in contact with regions of lower density, resulting in a wide range of Debye lengths: for such a simulation not to incur into stability issues, the grid spacing needs to be chosen to resolve the smallest Debye length in the system. 
Adaptive schemes can be devised to resolve in each region only the local Debye length and the local plasma frequency, resulting in considerable savings~\cite{fujimoto-sydora}: the use of adaptive grids resolving the \textsl{local} Debye length allows to save computational resources in the areas where high resolutions are not required.
The development of adaptive algorithms in the context of PIC methods is however complicated by the fact that PIC methods are essentially based on the assumption that the particles have a finite size which is fixed and equal to the cell size (see later Sec.~\ref{sec:particle_shape}), thus making the approach naturally suited to uniform grids and of non-straightforward application to non uniform grids. Nevertheless there are noteworthy attempts to expand the PIC method to non-uniform grids, as reviewed below for some of the most notable approaches of interest to space weather problems.\\ 
We organize such a review in three groups.
First, we consider Moving Mesh Adaptation (MMA) methods, where only one grid level is simulated (usually, adaptive grid methods prescribe the simulation of multiple grid levels), with a non-uniform structure and a connectivity that allows one to map the simulated grid to a uniform logical grid. 
The non uniform grid is regenerated at each time step as the end result of a process that distorts it by attracting more points in the regions of interest and away from regions where the system is smooth and uneventful. This approach has been followed for example within the Celeste Implicit Moment method and is reviewed in Ref.~\cite{jerrygrid1,jerrygrid2, lapenta_democritus}.\\
Second, Adaptive Mesh Refinement (AMR)~\cite{berger} can be used for the field part of the PIC method. Notable examples of this approach, which was developed especially for space weather application and has been shown to resolve the localized regions of reconnection embedded in larger systems~\cite{fujimoto-sydora}, are Ref.~\cite{vay-colella,fujimoto-sydora}.\\
Lastly, unstructured grids~\cite{brown-pic} can be used starting from finite element methods to discretize the fields.\\
All these approaches need to deal with the fact that, as the grid spacing changes locally, the particles and the cells cannot always have the same size. The PIC derivation is based on such an equality to provide a simple efficient interpolation between particles and cells (see, for example, Ref.~\cite{lapenta05} and Sec.~\ref{sec:particle_shape} later in the paper). The first two approaches relax the requirement that the particle size remains constant and base the interpolation on the local grid spacing: the particles have the local size of the cells they are embedded in. For the MMA approach this is achieved by assuming that the particle shape function is constant in the logical space $\bfxi$ rather than in the physical space $\bx$. If the shape functions are chosen in the logical space, the interpolation functions can still be computed with the b-spline chain rule and the same formulas of uniform grids can still be used. In the second group of adaptive techniques, a patch-based AMR approach is typically used: the particles have the size of the cells of the patch where they reside and interpolation can proceed in each patch as if it was a uniform grid.\\
This simplification of the interpolation step on adaptive PIC has a serious consequence: the exact derivation of Ref.~\cite{lapenta2012_spaceWeather} breaks down and the changing particle shape introduces new terms proportional to the temporal derivative of the shape function. The derivation of the PIC method outlined in Ref.~\cite{lapenta2012_spaceWeather} and briefly reminded later in the paper assumes a fixed particle shape: a simple analysis proves that, within this assumption, momentum is conserved exactly when the same interpolation function is used for all quantities~\cite{hockney,pic-russo} and provided that the solution of the Maxwell's equations also conserve momentum (i.e. the numerical Green's function reflects the properties of anti-symmetry of the exact continuum Green's function).
When the shape functions for particles change in time, instead, additional terms related to the temporal variation of the shape function have to be added to the equations of motion. However, in practice these additional terms are neglected and the resulting schemes do not conserve momentum, with the additional side effect that particles experience a self-force due to their traversing regions of different sizes and thus to the change in time of their shape function~\citep{vay-self,colella-self}. In the MMA approach, this error is controlled by using smooth grids where the spacing changes progressively: since the neglected terms are proportional to the changes in the shape function, when the changes are small the error is small. However, in the AMR approach the change in particle shape is sudden at the interface between patches and leads, first, to the necessity of refining the grids in small steps (refinement factors of two are the usual choice) and, second, to severe errors that need to be corrected with appropriately devised methods~\citep{colella-self,vay-self}.\\
In this paper, a novel approach which allows to avoid such inconveniences is presented.\\
The following two points are the main elements of discontinuity with respect to the existing implementations: first, while in the previous works the grid area needing an increased resolution was simulated completely with fields and particles only at the most refined level, here also the so-called "coarser grids" (i.e., the grids with larger spacial resolutions) are retained and fully simulated alongside the more refined areas, thus making this implementation a Multi Level Multi Domain (MLMD) approach. 
All the simulated domains are self-similar and fully functional, each of them being evolved in time for both fields and particles, which are bound to the birth level and whose shape function is tailored on the local grid size. Second, the baseline algorithm used here is the Implicit Moment PIC Method (see Sec.~\ref{sec:stability_solution}), in contrast with the explicit PIC used in the other cases. \\
The structure of the paper intends to guide the reader towards the motivations that lead to such an implementation.\\
First, in Sec.~\ref{sec:challenges}, the challenges to face in the development of adaptive PIC methods are recalled and pinpointed in the necessity to adapt the method to the variation of the grid size (Sec.~\ref{sec:particle_shape}) and in the stability dictated limits on the grid resolution required by explicit PIC methods (Sec.~\ref{sec:stability}). In Sec.~\ref{sec:solution} our solutions to such challenges are presented: using multiple, complete and self-similar domains (i.e., a MLMD approach) offers an elegant solution to the problem of adapting the particle shape functions to the changing grid resolution (Sec.~\ref{sec:particle_shape_solution}), while the Implicit Moment PIC algorithm allows to notably relax the above-mentioned stability conditions (Sec.~\ref{sec:stability_solution}). In Sec.~\ref{sec:overview} the approach proposed is presented in its conceptual guidelines, while in Sec.~\ref{sec:InterDomainInter} the details of the information exchange between the levels are provided. Sec.~\ref{sec:test} reports the results of a battery of tests performed in a 1D environment: a set of challenges is selected to address specific issues arising in a MLMD implementation and provides the occasion for a series of comments on fields and particles coupling in the system. Finally, conclusions are drawn in Sec.~\ref{sec:conclusions}.

\section{Challenges for adaptive Particle In Cell methods}
\label{sec:challenges}

In this Section the two main challenges to be faced when developing adaptive PIC methods are recalled. However, to fully appreciate them a reminder of the derivation of the PIC method is needed.\\
The PIC method is a mathematically rigorous discretization in a Lagrangian form of the Vlasov-Maxwell system describing a plasma at the kinetic level~\citep{hockney}. The phase space distribution function $f_s(\bfx,\bfv,t)$ for a given species $s$ (electrons or ions) is defined as the number density per unit element of the phase space and is governed by the Vlasov equation:
\begin{equation}
 \frac{\partial f_s}{\partial t}+\bfv\cdot \frac{\partial f_s}{\partial \bfx} +\frac{q_s}{m_s}(\bfE+\bfv \times \bfB) \cdot \frac{\partial f_s}{\partial \bfv}=0
\end{equation}
where $q_s$ and $m_s$ are the charge and mass of the species, respectively. The derivatives are written in the standard vector notation in the three-dimensional configuration space $\bfx$ and three-dimensional velocity space $\bfv$, forming together the classical phase space.

The electric and magnetic fields are given by the Maxwell's equations,
\begin{equation}
\begin{array}{c}
\label{eq:maxwell}
 \displaystyle \nabla \times \bfE=-\frac{\partial \bfB}{\partial t}\\ \\
\displaystyle \nabla \times \bfB=\mu_0 \epsilon_0 \frac{\partial \bfE}{\partial t}+\mu_0 \bfj \\ \\
 \displaystyle \epsilon_0 \nabla \cdot \bfE=\rho \\ \\
\displaystyle \nabla \cdot \bfB=0,
\end{array}
\end{equation}
where the sources of the Maxwell's equations are the first two moments of the particle distribution function, the 
 charge density,
\begin{equation}
 \rho(\bfx,t)=\sum_s q_s \int_{\mathbb{V}} f_s(\bfx,\bfv,t)d\bfv
\end{equation}
and the current density,
\begin{equation}
 \bfj(\bfx,t)=\sum_s q_s \int_{\mathbb{V}} \bfv f_s(\bfx,\bfv,t)d\bfv
\end{equation}
The integrations that define the moments are done over the velocity space ${\mathbb{V}}$.\\
The two key founding points of the PIC method that prevent its straightforward application in an adaptive grid framework are the following.\\
First, the size of the computational particles of the classical PIC method is taken to be constant in time and equal to the grid size. Transitioning to an adaptive grid would seem to require to sacrifice one of the two: either the particles have a constant shape that no longer fits the grid size as the particle moves in regions of different resolution or the particles size is always the local grid size, at the sacrifice of the constancy of the particle size. The constancy of the particle size is a cornerstone of the derivation and, if sacrificed, raises fundamental issues for particle motions, while the equality of particle and grid size is of great practical convenience in the implementation of the PIC method: giving it up means a much more cumbersome algorithm for grid-particle projection and interpolation. The approach presented here does not give up on either aspect and retains constant particle sizes while maintaining the size of particles and cells equal.\\ 
Second, PIC methods are strongly bound by stability limits: the grid spacing cannot exceed the local Debye length (with a factor of order 1) regardless of the scales of interest, thus significantly limiting grid adaptivity. This seemingly reduces adaptive PIC to methods to adapt the grid to the local Debye length, still a very useful tool in systems where the Debye length changes over a significant range, as it is indeed the case in many space and laboratory plasmas. However, this limitation should be removed for a truly adaptive PIC where the local grid spacing is chosen solely based on the accuracy of the resolution desired: consider that in many systems, in space and laboratory, quasineutrality is very strongly satisfied and nothing of interest happens at the Debye length. The method proposed bypasses the stability constraints of the explicit PIC method by adopting an Implicit Moment PIC method as the baseline algorithm to evolve in an adaptive direction.  
In the Sections below, the two apparent roadblocks mentioned above and the approaches proposed to overcome them are discussed in more details.

\subsection{Changing grid and particle size}
\label{sec:particle_shape}
In PIC methods, the phase space is discretized into a collection of super or computational particles, each representing a number of physical particles close to each other in  phase space. As a consequence of this assumption, the distribution function of each species $s$ is written as:
\begin{equation}
 f_s(\bfx,\bfv,t)=\sum_p f_p(\bfx,\bfv,t)
\label{defpart}
\end{equation}

where the index $p$ spans the computational particles of species $s$. The specific distribution function $f_{p}$ assigned to each computational particle has a number of free parameters which have the physical meaning of position and velocity of the computational particle and whose time evolution determines the numerical solution of the Vlasov equation. The distribution function $f_p(\bfx,\bfv,t)$ of a computational particle is thus assumed to be the following:

\begin{equation}
 f_p(\bfx,\bfv,t)=N_pS_\bfx(\bfx-\bfx_p(t))S_\bfv(\bfv-\bfv_p(t))
\label{deff}
\end{equation}
 where $S_\bfx$ and $S_\bfv$ are the {\it shape functions} in space and velocity of the computational particle and $N_p$ is the number of physical particles that are present in that element of phase space.
The common choices for the velocity and spatial shape functions are respectively Dirac deltas

\begin{equation}
 S_\bfv(\bfv-\bfv_p)=\delta(v_1-v_{1,p})\delta(v_2-v_{2,p})\delta(v_3-v_{3,p}),
\end{equation} 
 
and b-splines~\citep{splines}, usually of order 0 (Cloud In Cell, CIC methods),

\begin{equation}
 S_\bfx(\bfx-\bfx_p)=\frac{1}{\Delta x_{1,p} \Delta x_{2,p} \Delta x_{3,p}}b_l\left(\frac{x_1-x_{1,p}}{\Delta x_{1,p}}\right)b_l\left(\frac{x_2-x_{2,p}}{\Delta x_{2,p}}\right)b_l\left(\frac{x_3-x_{3,p}}{\Delta x_{3,p}}\right),
\label{shape-spline}
\end{equation}

where $\Delta x_{1,p}$, $\Delta x_{2,p}$ and $\Delta x_{3,p}$ are the lengths of the support of the computational particles in each spatial dimension.\\
When integrating the shape functions over the grid, i.e., when performing the following integral in all directions

\begin{equation}
\int_{\Delta x_i} S_x(x-x_p) d\bx =\int_{-\infty}^{\infty} S_x(x-x_p) b_0\left(\frac{x-x_g}{\Delta x_g}\right) d\bx,
\end{equation}

the interpolation functions $W(\bfx_g-\bfx_p)$ play a fundamental role. They are in fact used both to calculate particle moments on the grid points, as in

\begin{equation}
\begin{array}{l}
\displaystyle \rho_g=\frac{1}{V_g}\sum_p q_p W(\bfx_g-\bfx_p)\\ \\
\displaystyle \bfj_g=\frac{1}{V_g}\sum_p q_p \bfv_p W(\bfx_g-\bfx_p),
\label{eq:sources}
\end{array}
\end{equation}

where $V_g$ is the cell volume, and to calculate the fields acting on particles from the fields defined on the grid, as in

\begin{equation}
\begin{array}{l}
\displaystyle \bfE_p=\sum_g \bfE_g W(\bfx_g-\bfx_p)\\ \\
\displaystyle \bfB_p=\sum_g \bfB_g W(\bfx_g-\bfx_p).
\label{eq:fields_on_particles}
\end{array}
\end{equation}

In the cases when the support for the spatial shape function is equal to the grid spacing, $\Delta x_{i,p}=\Delta x_{i,g}$, and the shape function is a b-spline of order 0, the interpolation functions simply reduce to   

\begin{equation}
W(\bfx_g-\bfx_p)=b_{l+1}\left(\frac{x_g-x_p}{\Delta x}\right)b_{l+1}\left(\frac{y_g-y_p}{\Delta y}\right)b_{l+1}\left(\frac{z_g-z_p}{\Delta z}\right). \label{intfunction}
\end{equation}

As already mentioned, issues arise if $\Delta x_{i,p} \neq \Delta x_{i,g}$ (one of the possible solutions to the issue of particle shape in an adaptive grid, with particles of fixed shape moving across grids with different resolutions), since the mathematical description for the interpolation function becomes more complicated.\\
If the shape function does not explicitly include a time dependence, that is if 

\begin{equation}
 \frac{\partial S_{x_i}}{\partial t}=0, \label{nochangeshape}
\end{equation}

the equations for the motion of the computational particles in the PIC method follow directly from the Vlasov equation by simply noting that 
\begin{equation}
 \frac{d S_{x_i}(x_i-x_{i,p})}{d t}=\left.\frac{\partial S_{x_i}}{\partial {x_i}}\right|_{x_i=x_{i,p}}\cdot \frac{\partial x_{i,p}}{\partial t},
\end{equation}
where $x_i$ represents the three space variables. They are thus identical to those of a particle with center of mass properties given by the parameters of the computational particle, $\bfx_p$ and $\bfv_p$. If the terms $\partial S_{x_i} / \partial t$ are present, instead, as in the case of particle shapes being tailored on changing grid sizes (the most used solution in AMR PIC), the equations also include additional terms that make them not resemble anymore those of physical particles. In particular, they would include self-forces~\citep{vay-self,colella-self} that act on the superparticles in consequence of the change in the particle shape.  \\
One possibility to deal with the issue would be to limit the value of the terms $\partial S_{x_i}/\partial t$ with respect to the other terms of the equation of motion: if the particle shape changes slowly with the particle motion, the particle shape remains, in first approximation, an adiabatic invariant of motion. The equations for the superparticles remain then almost identical to those of real particles and the self forces become negligible. This approach, which requires the grid resolution to change slowly in space, has been followed in the variational grid adaptation ~\citep{jerrygrid1, lapentagrid} and has been implemented recently in the code DEMOCRITUS~\citep{lapenta_democritus}. The need to limit $\partial S_{x_i}/\partial t$ is incidentally one of the reasons why AMR PIC methods tend to prefer small resolution jumps between coarser and refined grid, the typical value being two.

Another possibility is to split particles that move from regions with lower resolution to regions with higher resolution, and, similarly, coalesce particles traveling in the opposite direction. Reliable methods have been developed to coalesce and split particles in order to force the particle population to maintain its total number per cell constant and have been successfully applied, for example, in Ref.~\cite{fujimoto-sydora}. The method does not attempt to exactly remove the self-forces created by the violation of the condition in Eq.~\ref{nochangeshape}, but rather eliminates them in a statistical, average way, compatibly with the statistical nature of the PIC approach. Note however that, while the splitting procedure for particles is relatively straightforward, as described in Ref.~\cite{lapentaadapt}, the inverse coalescence process is less immediate to code (particles with the same mass and charge and close in phase space have to be identified) and does not exactly conserve the total energy and the local distribution function~\cite{fujimoto-sydora}.\\
Finally, the exact removal of the self-forces can be achieved by giving up on their constant shape and complicating the interpolations and grid projection algorithms. This approach has been proposed within unstructured meshes methods~\citep{brown-pic} by simply computing the complex geometric overlap of a constant particle shape with the local grid arrangement. In AMR methods the grid is uniform in patches and the particle shape changes only at the boundary between regions of different resolution. There, the condition of Eq.~\ref{nochangeshape} is violated and a correction is applied.

\subsection{Stability dictated limits on grid resolution}
\label{sec:stability}

The stability dictated limits for explicit PIC methods and their consequences for plasma simulations have already been commented in Sec.~\ref{sec:intro} and~\ref{sec:challenges}. Let us recall here the origin of the necessity to resolve the fastest electron plasma frequency and the smallest Debye length in a system simulated with an explicit PIC method.\\
When an explicit formulation of the coupling between particles and fields is used, the particles are moved at each time step in the fields known at the beginning of the time step and the fields evolve for the same time step based on the recently updated and fixed particles. This works only if the time step is small, thus introducing the stability condition
\begin{equation}
\omega_{pe} \Delta t<2.
\end{equation}
The other stability condition, $c\Delta t<\Delta x$, stems from the explicit discretization of the Maxwell's equations and is normally not the main concern in space plasmas, since it can be easily removed with semi-implicit methods that discretize implicitly only the Maxwell's equations; the most challenging part is to keep the particle-field coupling implicit. 

A more serious drawback from the perspective of adaptive methods is that explicit PIC algorithms are subject to the finite grid instability. The mathematical study of the finite grid instability requires an analysis of all computational steps with the Laplace transformation of time and the Fourier transformation of space (see Ref.~\cite{birdsall,hockney} and citations therein); the summary of the analysis is that to avoid the finite grid instability in explicit particles methods, the grid spacing must be chosen to satisfy the constraint
\begin{equation}\label{finite-grid}
    \Delta x/\lambda_{De}<\varsigma
\end{equation}
where $\lambda_{De}$ is the Debye length and the parameter $\varsigma$ is of order one and depends on the details of the implementation. For the CIC scheme, the literature reports $\varsigma\approx\pi$.

The practical consequence of this numerical instability is a tremendous numerical heating of the plasma characterized by an alternatively positive-negative variation of the electric field accompanied by a correlated zig-zag perturbation of the phase space. Within a few cycles, the energy reaches the bounds of overflow in the machine representation of numbers. The finite grid instability must thus be avoided at all costs, requiring its stability constraint ($\omega_{pe}\Delta t<2$ and $\Delta x/\lambda_{De}<\varsigma$) to be respected everywhere in the domain for all directions for the fastest $\omega_{pe}$ and the smallest $\lambda_{De}$ in the system. Indeed, a practice advice is to use a considerably smaller time step of order $\omega_{pe} \Delta t=0.1$ and $ \Delta x< \lambda_{De}$, to avoid numerical heating~\cite{birdsall}.

It has already been observed that an AMR approach helps reducing the impact of the explicit PIC stability constraints in cases where the processes of interest are at the Debye length level but the presence of regions of higher density in contact with regions of lower density results in a large range of Debye lengths. However, less fortunate situations may reduce the benefits granted even by the application of a AMR algorithm to an explicit PIC. Consider, for example, that most reconnection events in space are dominated by processes that are on the electron inertial length, $d_e=c/\omega_{pe}$. This scale is still two orders of magnitude larger than the Debye length: having to resolve it purely for numerical stability reasons implies a waste of two orders of magnitude in each direction, a million times more cells than necessary~\citep{lapenta2012_spaceWeather, fujimoto06}.

If using uniform grids, this constraint thus requires to use cell sizes several orders of magnitude smaller than the scales of interest. However, also if adaptive grids are used the consequences of the finite grid instability on explicit PIC simulations are heavy, since the range within which the refinement factor of the refined grids can be chosen is severely limited. Indeed, the coarsest cells still need to resolve the local Debye length even in regions where there is no interest in having resolutions that small. Taking again the example of the regions in Fig.~\ref{scales}, to study a reconnection event in the magnetotail the ideal situation would be to resolve up to the electron skin depth in the vicinity of the reconnection region, but up to the ion skin depth or to even larger scales outside of it. Unfortunately, that is not possible even using adaptive explicit PIC.

\section{Tackling the challenges}
\label{sec:solution}

In this Section, the challenges to the development of PIC methods for adaptive grids are tackled and solutions are proposed. A MLMD approach is used to bypass the issue of the size of the particle shape functions in adaptive grids and results into introducing additional benefits to the system, such as the possibility of sudden jumps in the Refinement Factor and ease of programming (Sec.~\ref{sec:particle_shape_solution}). The strict stability conditions for the explicit PIC methods, so harmful also in the prospect of an adaptive grid evolution, are instead dealt with by switching to an Implicit Moment PIC approach (Sec.~\ref{sec:stability_solution}).  

\subsection{Multi Level Multi Domain approach}
\label{sec:particle_shape_solution}

The adaptive approaches conceptually closest to the present implementation~\citep{vay-self,fujimoto06} choose, for the sake of performances, to simulate computational particles only at the most refined levels and to assign them a shape function with dimensions equal to the most refined grid spacing. This solution raises the issue of self-forces and the need for corrective actions~\citep{colella-self}.\\
In this paper, a different, MLMD solution to the particle size dilemma is proposed: instead of relegating particles to the most refined levels, the system is simulated as a cloud of overlapping domains, each logically complete in both fields and particles and fully identical to the others, except in position, resolution and size.\\ 
Particles are created at initialization in each level with the size of the spatial shape function of Eq.~\ref{shape-spline} equal to the spacing of the grid they belong to and are bounded to their level of origin, not being allowed to transition from coarser to refined grids and vice versa. The details about how particles are dealt with at the boundaries between the coarser and refined grids are provided later in Sec.~\ref{sec:particleRepop}; for the purpose of the present discussion it is sufficient to mention that refined grid particles are lost when they exit their domain of origin and new refined level particles are created at the boundary cells of the refined grids with the splitting algorithm devised in Ref.~\cite{lapentaadapt} and already used in Ref.~\cite{fujimoto06}. Such an algorithm replaces a coarser grid particle with a collection of refined grid particles whose shape functions collectively reproduce the spatial extension of the original space function, thus eliminating self forces at the boundary in a statistical way. \\
This apparently simple MLMD solution grants a number of advantages. The fact that the shape function of particles has the same size of the grid granularity means that the simple interpolation functions of Eq.~\ref{intfunction} can be used to straightforwardly calculate the sources of Maxwell's equations (Eq.~\ref{eq:sources}) and the fields acting on a particle (Eq.~\ref{eq:fields_on_particles}): no complicated interpolation techniques need to be devised. More importantly, the particle shape function staying constant in time implies that no self forces are exerted over the particles, thus eliminating the need to correct for them. Lastly, since a refined area is simulated also with the coarser resolution both in fields and particles, no coarser grid particles have to be created by coalescence of the refined grid ones, thus avoiding the problems of conservation of total energy and local distribution function reported in Ref.~\cite{fujimoto-sydora}: only the "safest" part of the algorithm of Ref.~\cite{lapentaadapt} is used here. 

\begin{figure}[ht]
\centering
 \includegraphics[width=6cm]{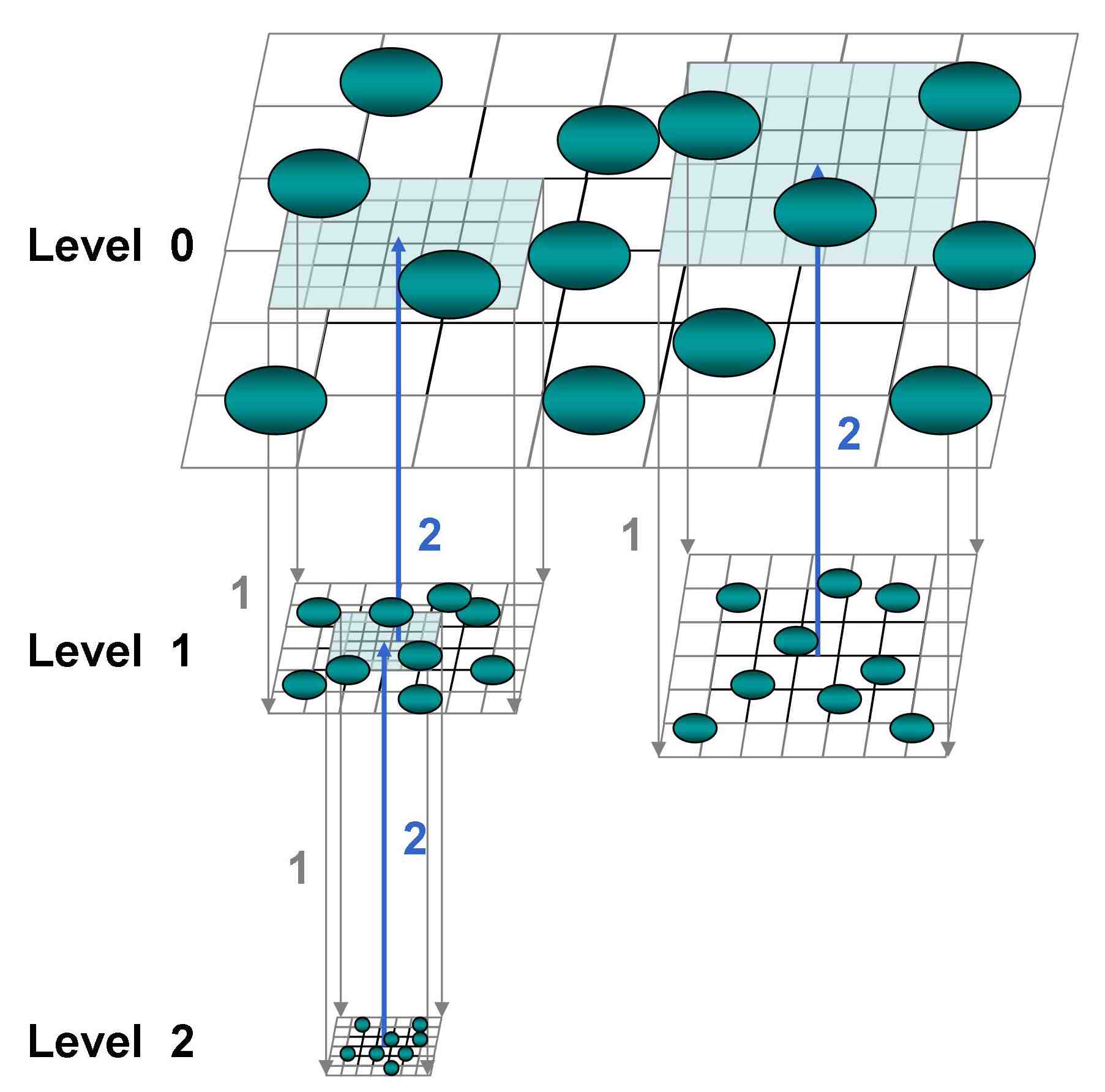}
\caption{Proposed Multi Level Multi Domain approach. The system is simulated as a cloud of self-similar, partly overlapping domains all complete in fields and particles. Particles are present at all levels, including the overlapping areas, and are sized according to the local grid size. Each level executes the same operations for fields and particles and the exchange of information for the fields is limited to the "downwards" interpolation of the boundary conditions for the refined levels from the coarser level fields (1) and the "upwards" projection of the refined fields from the refined to the coarser levels (2). Particles are regenerated at the boundaries of the refined grid with a splitting algorithm and are lost when they exit the refined areas, thus eliminating the need of coalescence operations.}
\label{cloud}
\end{figure}

A graphical depiction of the proposed system is given in Fig.~\ref{cloud}: an arbitrary number of levels with different Refinement Factors is fully simulated in fields and particles. Particles are present at each level with shape function sized on the local grid size and are used to calculate locally the sources of the Maxwell's equations, which are solved for $E^{n+\theta}$ on each level. Each level completes the same operations as the others and the communication between the grids is limited to the "downwards" (interpolation of the boundary conditions for the refined levels from the coarser level fields) and "upwards" (projection of the refined fields from the refined to the coarser levels) operations marked respectively with number (1) and (2) in the picture and described in more details in Sec.~\ref{sec:InterDomainInter}.

Such a system, apart for providing an elegant solution to the shape function problem, presents a few more aspects which made the MLMD approach attractive to the authors.

Unlike other AMR approaches, in fact, it removes the need to proceed in soft jumps of resolution differential. Typically AMR methods (see, for example, Ref.~\cite{fujimoto-sydora}) progressively create cells smaller by a factor of two in each direction, due to the increase of the self-forces with increasing difference between the particle shape function on the different levels and to the explicit PIC stability constraints. Here (also thanks to the implicit moment implementation, see Sec.~\ref{sec:stability_solution}), such a necessity is eliminated. This is a relevant achievement considering that kinetic problems tend to jump in big steps. Consider, for example, the now familiar magnetic reconnection scenario, with ions and electrons becoming unmagnetized at different distances from the X-line: in an ideal world, an adaptive method should allow a Refinement Factor proportional to the square root of the mass ratio in transitioning from the ion to the electron diffusion region. Moreover, the concept presented is intrinsically simple and object oriented in nature and as such the coding of the algorithms is by far easier than most AMR codes based on trees or patches. The promise of simplicity is not just in the conceptual ingredients but also in the operations, which are identically performed on all levels, regardless of the level of grid refinement. 

\subsection{The Implicit Moment Particle In Cell method}
\label{sec:stability_solution}

Implicit methods have been considered for several decades~\cite{mason1981,denavit1981,brackbill-forslund,directimplicit} an answer to the problem of the large number of time steps and the high resolution needed to run realistic problems with explicit particle methods. Here, an Implicit Moment PIC approach is adopted in the implementation typical of the family of codes started by Venus~\citep{brackbill82} and continued by Celeste~\citep{vu}, Parsek~\citep{parsek} and iPic3D~\citep{ipic}. The aim is to bypass the stability constraints of the explicit methods and allow more flexibility in the choice of the Refinement Factors.\\
In the presented implementation, the particle mover is based on the  $\theta$ scheme~\citep{vu,mover}:
\begin{equation}
\begin{array}{l}
\displaystyle \bfx_p^{n+1}=\bfx_p^{n}+\bfv^{n+1/2}_p \Delta t \\
\\
\displaystyle \bfv_p^{n+1}=\bfv_p^{n}+\frac{q_s\Delta
t}{m_s}\left(\bfE^{n+\theta}_p(\bfx^{n+1/2}_p)+\bfv^{n+1/2}_p\times
\bfB^{n}_p(\bfx^{n+1/2}_p)\right)
\end{array} \label{impl-mover}
\end{equation}
where a decentering parameter $\theta$, which is used to compute the weighted averages between the old and new time level quantities, as in $\Psi^{n+\theta}=\Psi^n(1-\theta)+\Psi^{n+1}\theta$, is used to vary the properties of the scheme.  

The velocity equation is more conveniently rewritten as:
\begin{equation}
\bv^{n+1/2}_p= \widehat{\bv}_p +\beta_s
\widehat{\bE}_p^{n+\theta}(\bx^{n+1/2}_p)
 \label{vel_eq}
\end{equation}
with $\beta_s=q_p \Delta t/2m_p$,
\begin{equation}
\begin{array}{c}
 \widehat{\bv}_p = \boldsymbol{\alpha}^n_s \cdot \bv^n_p \\ \\
\widehat{\bE}_s^{n+\theta} = \boldsymbol{\alpha}^n_s \cdot
\bE_s^{n+\theta},
\end{array}
\end{equation}
and $\boldsymbol{\alpha}_s^n$ defined as
\begin{equation}
\boldsymbol{\alpha}_s^n =  \frac{1}{1+(\beta_s B^{n})^2}
\left(\boldsymbol{I}-\beta_s \boldsymbol{I} \times \bB^n +\beta_s^2
\bB^n \bB^n \right)
\end{equation}
and representing a scaling and rotation of the velocity vector. 

\begin{figure}
\centering
\includegraphics[width=14cm]{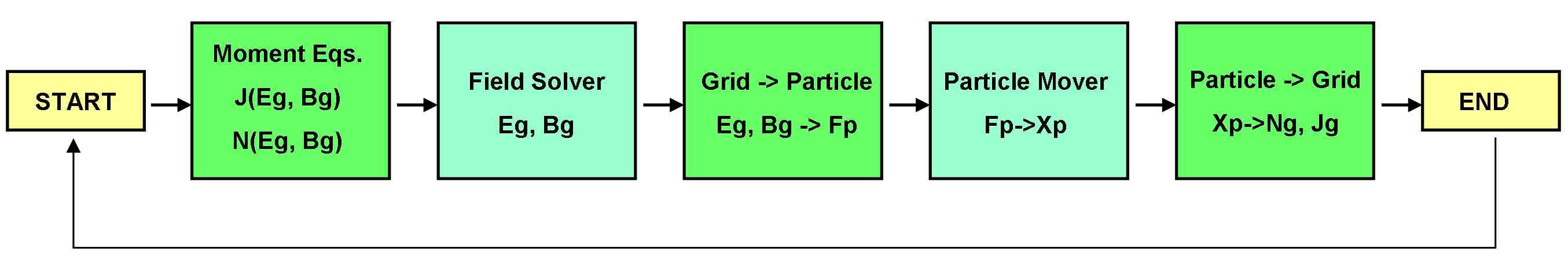}
\caption{Blocks needed in a cycle of the implicit moment PIC scheme. Modified from Ref.~\cite{lapenta05}.}
\label{implicitmoment}
\end{figure}

Eq.~\ref{impl-mover} and Eq.~\ref{vel_eq} for particle motion are in implicit form. 
In the Implicit Moment method, the Maxwell's equations, discretized as in
\begin{equation}
\begin{array}{c}
\displaystyle \mathbf{B}_g^{n+1}-
\mathbf{B}_g^n =-\Delta t \nabla \times \mathbf{E}_g^{n+\theta}  \\ \\
\displaystyle
 \mathbf{E}_g^{n+1}-\mathbf{E}_g^n
    =   \frac{\Delta t}{\mu_0\epsilon_0}\left(\nabla \times \mathbf{B}_g^{n+\theta} - \mu_0\mathbf{j}_g^{n+\frac{1}{2}}\right) \\ \\
\displaystyle  \epsilon_0 \nabla \cdot   \mathbf{E}_g^{n+\theta}
= \rho_g^{n+\theta} \\ \\
  \nabla \cdot \mathbf{B}_g^{n+1}  = 0,
\end{array}\label{impl-solver}
\end{equation}
are also solved implicitly, introducing a coupling between the equations for particle motion and those for the fields. This dependence is removed in the Implicit Moment method as summarized in Fig.~\ref{implicitmoment}, modified from Ref.~\cite{lapenta05}, and described in Refs. \cite{brackbill-forslund,vu,lapenta05,directimplicit, noguchi}: the sources of the field equations are approximated using the moment equations instead of being calculated directly from the particles. 

Once the field equations are solved within this approximation, the rest of the steps can then be completed sequentially: with the new fields, the particle equations of motion can be solved (either with a predictor corrector PC~\citep{mover} or a Newton approach~\cite{kelley}) and the new current and density can be computed for the next computational cycle.

The key step is thus to derive a suitable set of moment equations that can approximate the response of the plasma particles to the fields over a computational cycle. This is done through a series expansion of the interpolation function appearing in the expression for the field sources. The approximation is performed with respect to the particle position, choosing as the center of the expansion the particle positions at the beginning of the computational cycle (in direct implicit methods, instead, the expansion is centered around a guess of the new advanced position~\citep{directimplicit}):

\begin{equation}
W({\bf x}-{\bf x}_p^{n+1}) =W({\bf x}-{\bf x}_p^{n})+({\bf x}-{\bf
x}_p^{n}) \cdot \nabla W({\bf x}-{\bf x}_p^{n})+\frac{1}{2} ({\bf
x}-{\bf x}_p^{n}) ({\bf x}-{\bf x}_p^{n}) :\nabla \nabla W({\bf x}-{\bf
x}_p^{n})+ \ldots \label{expansion}
\end{equation}
where a tensor notation is used. 

The expansion of Eq.~\ref{expansion} can be used to compute the field sources directly using particle information only from  the previous computational cycle and removing the need to iterate over the particle and field equations. The details of the simple but tedious algebraic manipulations are provided in Ref.~\cite{vu}, the final answer being:
\begin{equation}
\begin{array}{c}
\rho_s^{n+1}=\rho_s^{n}- \Delta t \nabla \cdot \bfj_s^{n+1/2} \\ \\
\bfj^{n+1/2}_s=\widehat{\bfj}_s-\frac{\Delta t}{2}\boldsymbol{\mu}_s
\cdot E^{n+\theta} -\frac{\Delta t}{2} \nabla \cdot
\widehat{\boldsymbol{\Pi}}_s \label{momentimplicit}
\end{array}
\end{equation}
where the following expressions were defined:
\begin{equation}
\begin{array}{c}
\widehat{\bfj}_s = \sum_p q_p  \widehat{\bv}_p W({\bf x}-{\bf
x}_p^{n})\\ \\ \widehat{\boldsymbol{\Pi}}_s =\sum_p q_p
\widehat{\bv}_p
\widehat{\bv}_p W({\bf x}-{\bf x}_p^{n}) \\ \\
\end{array}
\end{equation}
with the meaning, respectively, of current and pressure tensor based on the transformed hatted velocities. The $\boldsymbol{\mu}_s$ term is an effective dielectric tensor which expresses the feedback of the electric field on the plasma current and density:
\begin{equation}
\boldsymbol{\mu}_s^n = - \frac{q_s \rho_s^{n}}{m_s}
\boldsymbol{\alpha}_s^n.
\end{equation}

The expression in Eq.~\ref{momentimplicit} for the sources of the Maxwell's equations provide a closure for the Maxwell's equations themselves. When Eq.~\ref{momentimplicit} is inserted in Eq.~\ref{impl-solver}, the Maxwell's equations can be solved without further coupling with the particle equations, as in the consistent second order formulation from Ref.~\cite{riccijcp}: 

\begin{equation} \label{equiv tottot}
\begin{array}{l}
\displaystyle{\left( c \theta \Delta t \right)^2 \left[ - \nabla^2
\mathbf{E}^{n+\theta} - \nabla \nabla \cdot \left(
\boldsymbol{\mu}^n \cdot \mathbf{E}^{n+\theta} \right) \right] +
\boldsymbol{\epsilon}^n \cdot \mathbf{E}^{n+\theta} } \cr
\displaystyle{= \mathbf{E}^n+ \left( c \theta \Delta t \right)
\left(\nabla \times \mathbf{B}^n-\frac{4\pi}{c}
 \mathbf{\hat{\bfj}}^n \right)
- \left( c \theta \Delta t \right)^2 \nabla 4 \pi \hat{\rho}^n},
\end{array}
\end{equation}
where $\boldsymbol{\mu}^n=\sum_s \boldsymbol{\mu}_s^n$ and
$\boldsymbol{\epsilon}^n={\bf I}+\boldsymbol{\mu}^n$, $\bf I$ being
 the identity tensor.

As shown in Ref.~\cite{riccijcp}, the second order formulation for the electric field needs to be coupled with a divergence cleaning step to ensure that Gauss's law (Eq.~\ref{eq:maxwell}-3, discretized in Eq.~\ref{impl-solver}-3) holds true at each time step.
The magnetic field is then computed directly once the new electric field is known, as in Eq.~\ref{impl-solver}-1.
The discretized equations and their boundary conditions form a non-symmetric linear system that is solved using the Generalized Minimal Residual (GMRES)~\citep{saad} method. For the divergence cleaning equation, Conjugate Gradient (CG) methods or Fast Fourier Transform (FFT) methods on uniform grids~\citep{saad} can be used since the discretized equation leads to a symmetric matrix.

The above summarized Implicit Moment formulation has the notable advantage of removing the strict stability limits of the explicit PIC.\\ The stability properties of the implicit moment method have been studied extensively in the past~\cite{brackbill-forslund}: the implicit  particle mover removes the need to resolve the electron plasma frequency and the implicit formulation of the field equations removes the need to resolve the speed of light. The time step constraints are thus replaced by an accuracy limit arising from the derivation of the moment equations using the series expansion in Eq.~\ref{expansion}. This limit restricts the mean particle motion to one grid cell per time step~\cite{brackbill-forslund}, i.e.
\begin{equation}
\label{time step limit 1} v_{th,e} \Delta t/\Delta x < 1,
\end{equation}
The finite grid instability limit for the explicit method, $\Delta x
< \varsigma \lambda_{De}$, is replaced by~\cite{brackbill-forslund}
\begin{equation}
\Delta x < \varepsilon^{-1} v_{th,e} \Delta t ,
\end{equation}
that allows large cell sizes to be used when large time steps are taken.

This is the key feature of interest of the implicit method: time and space scales can be chosen freely according to the desired accuracy (refer to Refs.~\cite{birdsall,brackbill1985,brackbill-forslund} for details on how the non-resolved scales are dealt with), but their ratio is not free, since it must stay within the bounds just outlined:
\begin{equation}
\varepsilon< v_{th,e} \Delta t/\Delta x < 1,
\label{impl-stab}
\end{equation}
The upper limit is generally the main concern and should never be violated. The lower limit, $\varepsilon$, is in practical cases usually a very small number that can often be approximated by zero. 

The gain afforded by the relaxation of the stability limits is already notable in non-adaptive cases, but becomes particularly precious in the adaptive case.

As regards the non adaptive cases, the new stability constraints remove the necessity to resolve the faster electron plasma frequency and the smaller Debye length while retaining the sub-$\Delta t$ scales and thus the possibility to exchange energy between sub-$\Delta t$ fluctuations and particles (notice that in other approaches, e.g. the gyrokinetic or hybrid one~\citep{multiscale}, such processes are completely removed and the energy channel towards them is interrupted). Instead, when adaptive schemes are considered the increased range of time steps and grid spacings available means that the level of refinement between coarser and refined levels may be chosen more freely, without the concern of resolving the Debye length in all grids. As regards the choice of the time steps, two possibilities are now open: the time step can either be adjusted to the grid spatial resolution, with the refined levels using a fraction of the time step of the coarser levels in order to maintain the same $\Delta x/ \Delta t$ ratio across the levels, or $\Delta t$ can be set for the entire system to the value required by the more refined level, the critical part of the inequality of Eq.~\ref{impl-stab}. Here this second solution is preferred, since in a future parallel implementation with exchange of information between the levels adopting a larger time step for the coarser domains in absence of a sophisticated work sharing algorithm would just make the coarser grids processors remain idle while waiting for communication from the refined grids processors. Notice again that the $\varepsilon< v_{th,e} \Delta t/\Delta x$ part of the stability constraints is, according to practical experience, not a significant hindrance and that the value of $\epsilon$ can be rather small, thus not significantly limiting the spacial grid resolution of the coarser grids when the $\Delta t$ is chosen according to the refined grids requirements.\\
Observe additionally that Implicit Moment methods come with an embedded numerically-enhanced Landau damping at high wavenumbers. This characteristic proves to be a precious ally against the issue of the reflection of low-$\lambda$ waves, supported by the refined grids but not by the coarser grids, at the boundaries between the levels~\citep{vay-self}, a problem which becomes dramatic for explicit algorithms if certain stencils are used for the discretization of the fields~\citep{fujimoto11}.

\section{Overview of the Multi Level Multi Domain method}
\label{sec:overview}
To summarize the argumentations developed above, the MLMD approach proposed here is based on the following principles:
\begin{itemize}
\item the physical system is represented as a cloud of overlapping domains organized in levels, each one of them conceptually identical to the others except for a scale factor and a translation. At the coarsest (upper level in Fig.~\ref{cloud}), the whole domain is discretized in a coarse uniform grid. At the next levels, the areas of interest are further discretized by other domains identical to the upper levels in everything (including number of cells and particles) except for being rescaled in size down to a fraction, not necessarily in soft jumps;
\item  while the top coarsest level covers the whole computational domain, the lower levels with higher resolution cover only progressively smaller parts of the physical system, the areas where the enhanced resolution is really needed;
\item the scaling affects cells and particles alike, with particles populating all the levels, not just the most refined, and having the same size as the cells they move in. As a consequence, the particles of the refined levels have a smaller shape function than those of the coarser levels;
\item  since each level has its own cells and particles and performs the same operations as the others, from a software point of view all levels are identical and in a object oriented approach they are different instantiations of the same class. For each level, the instantiation is rescaled and shifted;
\item the communication between the levels consists of the projection of the refined level fields to the coarser levels and the interpolation of the boundary conditions for the refined levels from the coarser level information. Particles at the boundary of the refined levels are regenerated according to the particle distribution at the coarser level. The details about inter-domain communications are provided in Sec.~\ref{sec:InterDomainInter}; 
\item the method used to advance particles and fields on each domain is the Implicit Moment PIC method.

\end{itemize}

\begin{figure}[ht]
\centering
 \includegraphics[width=14cm]{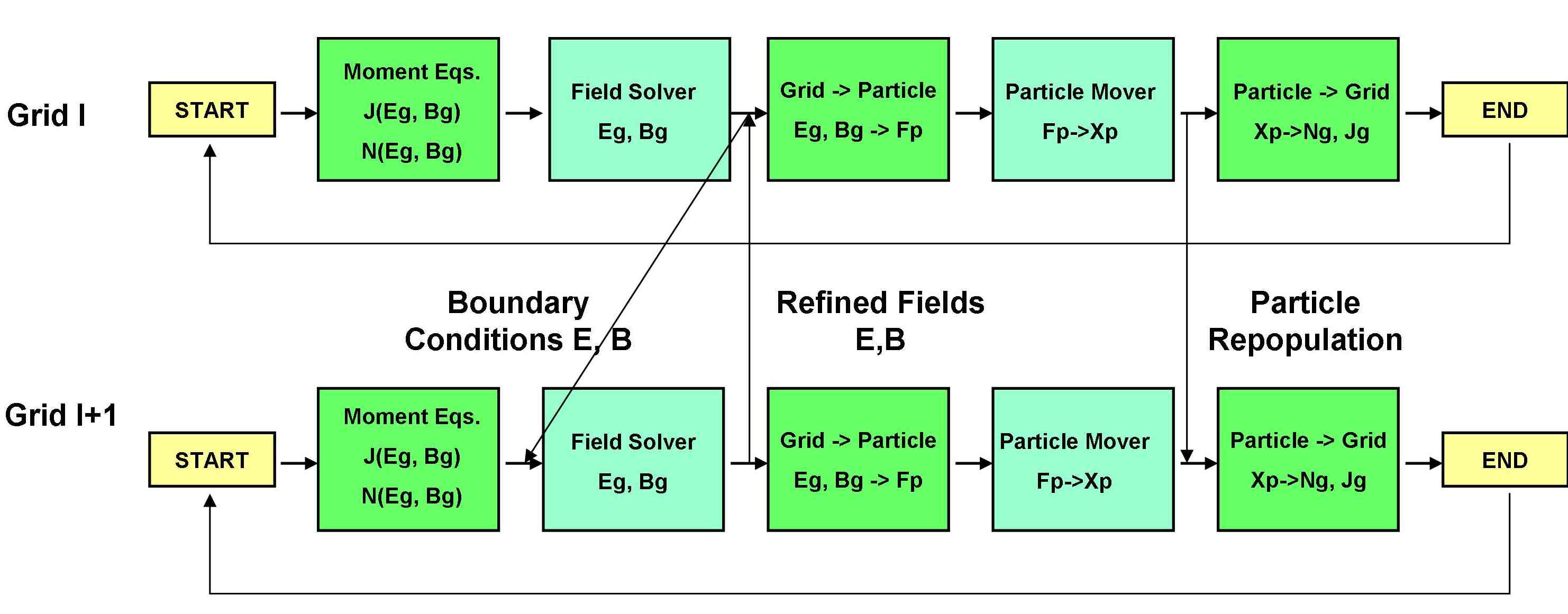}
\caption{Representation of one computational cycle advancing particles and fields on each domain of the cloud. Each domain is completing the same operations as all the others, with three points along the pipeline of operations (marked by vertical arrows) where communications across the levels become necessary. The fields from the coarser grids are used to provide boundary conditions for the refined grid fields ("Boundary Conditions E, B") before the field solver starts at the refined grid levels. After the field solution is known on the refined grids, it is projected to the coarser domains ("Refined Fields E, B"). After the particle mover step and before calculating particle moments at the refined grid level, the refined particles at the boundaries are regenerated according to the coarser grids particle distributions ("Particle repopulation").}
\label{collective}
\end{figure}
The computational cycle involving collectively all the cloud of domains can then be represented as in Fig.~\ref{collective}, which depicts the modes and timing of the inter-level communications superimposed to the Implicit Moment method cycle of Fig.~\ref{implicitmoment}. \\
The computation starts on each domain with the moments equations being computed from the particle moments previously calculated. At this point, field information from the coarser levels is needed to provide the boundary conditions for the electric and magnetic fields on the refined grids ("Boundary Conditions E, B" step in the picture). After computation of the fields at time $n+1$, the refined grid fields are projected to the coarser levels ("Refined Fields E, B"), thus updating their values with the increased level of accuracy granted by the refined grids. Then the computation resumes on each domain independently, with the interpolation of the field data to the particles, followed by the movement of the particles in the newly computed fields. When the new particle data are available at the coarser levels, particles are repopulated at the refined levels according to the parent distributions ("Particle repopulation"), thus enabling the calculation of particle moments on all levels. Notice that the repopulation of particles at the refined levels must be done before the calculation of particle moments, since the already updated density must be used to enforce Gauss's law before advancing the fields at the next time step.

\section{Inter-domain interaction}
\label{sec:InterDomainInter}

In this Section, the inter-domain communications represented by vertical lines in Fig.~\ref{collective} are explained in details. Three communication steps are needed across the grids to perform the following operations:
\begin{itemize}
\item project the fields from the refined to the coarser levels (Sec.~\ref{sec:upwards});
\item interpolate the boundary conditions for the refined levels electric and magnetic fields from the coarser levels fields (Sec.~\ref{sec:downwards});
\item generate the refined level boundary particles from the coarser levels particle distribution. Particles can leave the domain in the refined levels, but similarly new particles, created using information from the coarser levels, should be created in the boundary cells with positions and velocities derived from the coarser level particle distribution. Notice that particle information is exchanged only from the coarser to the refined grids (no particle coalescence is needed, since particles are simulated also at the coarser grid levels) and that the particle moments (densities, currents, pressures) in the coarser grids are calculated natively from the local particles and not projected from the refined grids.
\end{itemize}

\subsection{Projection from refined to coarser levels}
\label{sec:upwards}

Information is transferred from the refined to the coarser grids through the "upwards" projection of the electric field and the consequent re-calculation of the magnetic field.\\
The three components of the electric field are projected from the refined to the coarser grids as in 
 
\begin{equation}
\bfE_{P,g_l}=\frac{1}{2}\left(\bfE_{N,g_{l}}+\frac{\sum_{g_{l+1}}\bfE_{N,g_{l+1}}W_{g_{l}}(\bfx_{g_{l}}-\bfx_{g_{l+1}})}{\sum_{g_{l+1}}W_{g_{l}}(\bfx_{g_{l}}-\bfx_{g_{l+1}})}\right),
\label{eq:proj_E}
\end{equation}

where the subscripts $g_{l}$ and $g_{l+1}$ refer to the coarser and refined grid respectively, the coarsest grid being labeled with the index $0$, $\bfE_{P}$ denotes the projected electric field and $\bfE_{N}$ the electric field calculated natively on the grid.
Notice that $\bfE_{P,g_{l}}$ on the coarser grid is obtained as the average of two components: the electric field $\bfE_{N,g_{l}}$ calculated natively on the coarser grid and the projection to the level $g_{l}$ of the electric field calculated at the level $g_{l+1}$. Such projection is done as if each node in the refined grids was a particle in the coarser grids, with a normalizing factor as denominator: $W_{g_{l}}$ is the same interpolation function defined in Eq.~\ref{intfunction} and used in Eq.~\ref{eq:sources} to obtain the particle sources on the grid points starting from the positions and velocities of the particles.\\
The projection method of Eq.~\ref{eq:proj_E} is a key difference between the MLMD approach presented here and the more common AMR algorithms which discard the information from the coarser levels: here the fact that both the coarser and the refined levels are fully functional allows to combine the information from both instead of simply neglecting the coarser grid information. \\
Notice that a similar approach has been adopted, albeit only in a limited portion of the coarse domain, in Ref.~\cite{sugiyama07}, where a coarse grid simulated with Magneto-Hydro-Dynamic (MHD) equations is interlocked with a refined PIC domain, as part of a strategy to eliminate high-frequency noise injected from the refined to the coarser domain through the upwards projection of the fields. Indeed, it is similarly observed here that calculating $\bfE_{P,g_{l}}$ as an average instead of simply substituting it is beneficial to the evolution of the MLMD system in cases where the growth of a strong longitudinal electric field is not expected (e.g.: the Maxwellian, Weibel and shock benchmarks described in Sec.~\ref{sec:test}). In fact, when the projected electric field is calculated by substitution, the particle density $n_{g_{l}}$ and the native field $\bfE_{N,g_{l}}$ in the coarser grid area which overlaps the refined grid develop a significantly higher level of noise when compared to the neighboring cells, thus introducing the risk of using non optimal electric field values as boundary conditions for the refined grids. It is easy to understand why a noisier density $n_{g_{l}}$ degrades $\bfE^{n+\theta}_{N,g_{l}}$ in the area of overlap: the divergence cleaning step~\cite{riccijcp} uses a degraded $n_{g_{l}}$ to "correct" $\bfE^{n}_{P,g_{l}}$, which is later used to compute the Right Hand Side of Eq.~\ref{equiv tottot} and calculate the native field $\bfE^{n+\theta}_{N,g_{l}}$ at time $n+\theta$.\\ It is less straightforward to understand why, when projecting the electric field by substitution, the particle density $n_{g_{l}}$ is noisier in the overlap area, since particles are moved in the projected and not in the native fields. A possible explanation is that, when the simulated problem does not cause the growth of a strong longitudinal field, the longitudinal field on the coarser and refined levels reduces to noise fluctuations whose sign and amplitude may be different for the coarser and refined levels at a corresponding coordinate. This may introduce spurious effects in the coarser level particle motion, since particles are moved with the native coarser grid field in the area of no overlap and with the projected refined grid field in the overlap area: consistency issues may arise at the boundary area, when particles are subject to the abrupt change from the native to the projected field and vice versa, and be amplified during the following steps of the calculation.\\
When the electric field projection is done by averaging, instead, the particle density and the native electric field in the overlap area exhibit a level of noise comparable to the one in the neighboring areas. This is due to the fact that half of the native field information is retained when calculating the projected field and thus an effective coupling between the refined and coarser grid is reached, allowing particles to experience a smoother field transition when moving across the overlap area.\\
The magnetic field in the overlap area of the coarser grids is calculated according to the discretized Maxwell-Faraday equation (Eq.~\ref{impl-solver}-1) from the projected electric field:

\begin{equation}
\displaystyle \mathbf{B}_{P,g_{l}}^{n+1}= \mathbf{B}_{P,g_{l}}^n -\Delta t \nabla \times \mathbf{E}_{P,g_{l}}^{n+\theta}  \\ \\
\label{eq:proj_B}
\end{equation}
The calculation of the magnetic field through Eq.~\ref{eq:proj_B} is preferred to the most immediate alternative, an average between the coarser and refined grid fields similar to Eq.~\ref{eq:proj_E}, since the Maxwell-Faraday equation at grid level is considered paramount to preserve.\\ 
As regards the timing of the projection operations, it is important to remind (see also Fig.~\ref{collective}) that they are carried out \textit{before} moving particles in the coarser grid in order to increase the consistency between projected fields and particles which, as it was argumented before, is fundamental in the MLMD system.
The base equations for particle motion on the coarser grids, originally expressed as in Eq.~\ref{impl-mover}, thus become

\begin{equation}
\begin{array}{l}
\displaystyle \bfx_p^{n+1}=\bfx_p^{n}+\bfv^{n+1/2}_p \Delta t \\
\\
\displaystyle \bfv_p^{n+1}=\bfv_p^{n}+\frac{q_s\Delta
t}{m_s}\left(\bfE^{n+\theta}_{Pg_{l}}(\bfx^{n+1/2}_p)+\bfv^{n+1/2}_p\times
\bfB^{n}_{Pg_{l}}(\bfx^{n+1/2}_p)\right),
\end{array} 
\label{impl-mover-AMR}
\end{equation}

where the projected fields $\bfE^{n+\theta}_{Pg_{l}}$ and $\bfB^{n}_{Pg_{l}}$ are used instead of the native ones in a Predictor Corrector approach. The finest grid particles are moved, of course, in the native fields.

\subsection{Interpolation from coarser to refined levels}
\label{sec:downwards}

In order to drive the evolution of the refined levels, their boundary points are obtained by interpolation from the fields of the level above. Again the grid points of the refined levels are treated by the coarser levels as if they were particles and the interpolation is therefore:

\begin{equation}
\psi_{I,g_{l+1}}=\sum_{g_{l}}\psi_{N,g_{l}}W_{g_{l}}(\bfx_{g_{l}}-\bfx_{g_{l+1}})
\label{eq:BC_interp}
\end{equation}
where $\psi_{I,g_{l+1}}$ denotes the interpolated point on the refined grid and $\psi$ is a generic notation for both the electric and magnetic field. Notice that the boundary conditions for the refined grid are calculated from the native, not projected, coarser grid fields $\psi_{N,g_{l}}$ and that the points needing interpolation are the boundary nodes, e.g., the ghost cells nodes. This minimal information exchange for fields, limited to the ghost cells nodes, is sufficient to drive the refined grid evolution also in challenging cases as the shock benchmark presented in Sec.~\ref{sec:shock}.\\
Since the boundary conditions for the electric field are needed at the refined grid level \textit{before} calculating $\bfE_{N,g_{l+1}}^{n+\theta}$, a bottleneck for a future, planned parallel evolution of the presented algorithm is here foreseeable: the refined grids have to wait for the calculation of the coarser grid $\bfE_{N,g_{l}}^{n+\theta}$ before starting their own calculation. This bottleneck can be overcome by using as boundary conditions for the refined grid not the final result of the iterative calculation of Eq.~\ref{equiv tottot} on the coarser grid, but an intermediate one. The assessment of the viability of this solution is left as future research.

\subsection{Particle repopulation in the refined levels}
\label{sec:particleRepop}

In the presented MLMD approach, particles information has to be exchanged between the coarser and the refined grids only in the "downwards" direction: since the coarser levels are fully simulated, particles exiting the refined domains can be simply lost, without the need of using them to recreate particles on the coarser grids. However, refined grid particles have to be recreated at the boundaries of the grids consistently with the coarser grids particle distribution.\\
Two possibilities are open for particle repopulation: particles can either be recreated according to a fixed distribution function, e.g. a Maxwellian distribution function, with parameters (drift velocity, thermal spread) derived from the coarser grid, as in Ref.~\cite{sugiyama07} and Ref.~\cite{shay07} (notice however that in both these cases the coarser level is not simulated with a particle method and therefore full particle information is not available), or by taking full advantage of the fact that the velocities and positions of coarser grid particles are known, as in Ref.~\cite{fujimoto-sydora}.\\
In taking this decision, two factors have been considered: as a first attempt, we decided to retain the possibility to simulate non Maxwellian distribution functions at the grid interface, option which is lost if the first approach is adopted~\citep{shay07}. Secondly, we considered that particle repopulation is a task which requires particular attention since consistency in particle motion between the refined and coarser grids is not enforced explicitly in the MLMD algorithm presented here, but comes as a consequence of the projection and interpolation operations described in Sec.~\ref{sec:upwards} and Sec.~\ref{sec:downwards}.\\ For this reason, a rather conservative approach has been opted for in regenerating particles in order to guarantee the consistency between the particle population on the refined and coarser grids at least at the grid boundaries (further considerations about particle motion consistency across the grids will follow in Sec.~\ref{sec:2streams}). \\
First of all, particles are regenerated not only in the ghost cells, but also in a small number of neighboring cells already inside the active part of the refined grid. The total area where particles are regenerated is called here "Particle Repopulation Area" (PRA). This operation, which may seem redundant, proves instead to be of fundamental importance in cases when boundary particles are subject to the formation of structures in phase space, as in the Weibel and two stream instability benchmarks shown in Sec.~\ref{sec:test}. The particles populating the PRA from the previous time step are deleted and new particles are generated by reproducing the parent particle distribution in the corresponding area of the parent grid. Each particle in the parent grid sitting in an interval corresponding to the $PRA\pm dx_{g_{l}}$, where $dx_{g_{l}}$ is the coarser grid spacing, is split, on the refined grid, into $RF$ particles ($RF$ being the Refinement Factor, that is the ratio between the coarser and the refined grid dimensions) following the algorithm described in Ref.~\cite{lapentaadapt}, used in Ref.~\cite{fujimoto06} and summarized here in Eq.~\ref{eq:particleGenCharge}-\ref{eq:particleGenPos} for the $1D$ case (the extension to higher dimensions is straightforward, just requiring to adapt Eq.~\ref{eq:particleGenPos} in order for the combined shape function of the refined grid particles to reproduce the coarser grid particle shape function): 

\begin{equation}
\begin{array}{c}
\label{eq:particleGenCharge}

q^{n+1}_{p_{g_{l+1}}}= q^{n+1}_{p_{g_{l}}}/RF 
\end{array}
\end{equation}

\begin{equation}
\begin{array}{c}
\label{eq:particleGenVel}
\bv^{n+1}_{p_{g_{l+1}}}= \bv^{n+1}_{p_{g_{l}}}\\
 
\end{array}
\end{equation}

\begin{equation}
\begin{array}{c}
\label{eq:particleGenPos}

x^{n+1}_{p_{g_{l+1}},i}= x^{n+1}_{p_{g_{l}}}-\frac{dx_{g_{l}}}{2}+ dx_{g_{l+1}} \left( \frac{1}{2}+i \right)-x_{0,l+1},\; for\; i\,=\,0:RF-1  
\end{array}
\end{equation}
Eq.~\ref{eq:particleGenCharge} guarantees that the total charge is respected, Eq.~\ref{eq:particleGenVel} that the velocity distribution is reproduced without distortions and Eq.~\ref{eq:particleGenPos}, where $x_{0,l+1}$ is the coordinate referred to the coarser grid at which the active part of the refined grid starts, that the combined shape function of all the $RF$ refined grid particles corresponds to the shape function of the coarser grid parent particle.\\
Notice that, as visible in Fig.~\ref{collective} and remarked by the temporal index $n+1$ in Eq.~\ref{eq:particleGenCharge}-\ref{eq:particleGenPos}, the particles in the refined grid are generated from the parent grid distribution after the particle mover operations have been completed on both grids. The aim is to calculate particle moments consistent with the repopulated particle distribution on the refined grids.\\
As regards the number of cells to repopulate, that is, the size of the PRA area, the rule of thumb followed in Sec.~\ref{sec:test} is to dimension the repopulation area according to the distance that the particle with the maximum reasonably expected velocity in the system can cover in a time step. Future implementations will make the choice of the number of PRA cells automatic.

\section{Benchmarks}
\label{sec:test}

The simulation of a MLMD PIC system entails a variety of challenges, which span from checking fields and particle continuity at the grid boundaries to the more delicate tasks of assessing the impact of the exchange of information between the levels, verifying the results of the simulations against theoretical expectations and understanding which degree of consistency in the evolution of the different grid levels should be expected.\\ These issues are addressed here through a series of simulation challenges in a 1D environment, each devoted to test a specific aspect of the system:
\begin{itemize}
	\item {how the system reacts to the exchange of information between the levels. A Maxwellian plasma is simulated in a MLMD environment to examine how the two coupled grids react to the exchange of information in absence of notable plasma activity;}
	 \item {how an instability is simulated across two levels and how not moving particle structures are created and preserved across the grid boundaries. A Weibel instability is excited in the coarser and refined grids and its evolution is followed. In particular, the growth rate of the instability and the continuity of fields and particles across the grids are checked, with particular attention to the development of electron structures in phase space;}
	 \item {how moving particle structures react to crossing the grid boundaries. A two stream instability is simulated and the continuity of electron holes moving across the coarser-refined grid boundary is checked. Moreover, the comparison of the electric field and electron holes evolution of the two grids will stir considerations about particle motion consistency in the two grids;}
	 \item {how the refined grid reacts to strong driving from the coarser grid. A shock is excited at the coarser grid boundaries and launched towards the center of the coarser grid, where the refined grid sits. The propagation of the shock wave across the domains is studied.}
\end{itemize}
 
\subsection{Testing the system reaction to the exchange of information between the refined and coarser grids: Maxwellian plasma simulation}
\label{sec:max}
\begin{figure}[ht]
\centering
 \includegraphics[width=8cm]{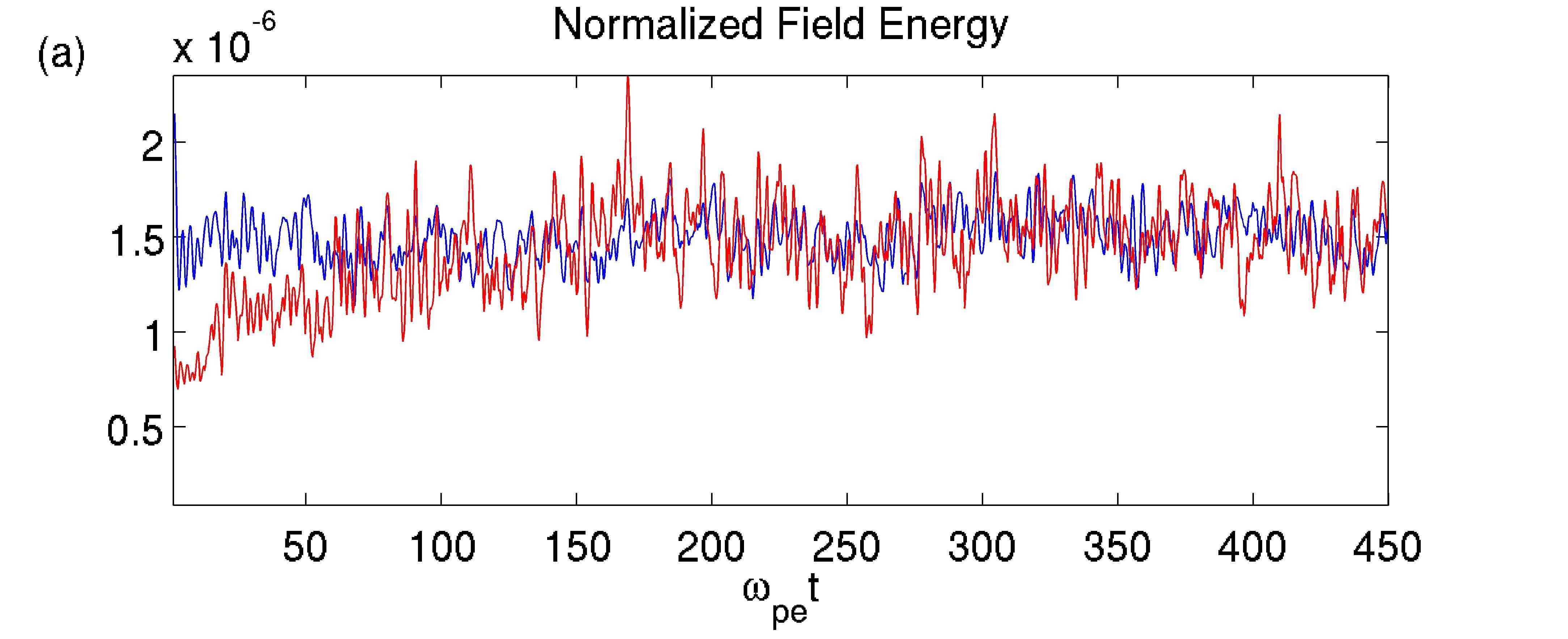}
 \includegraphics[width=8cm]{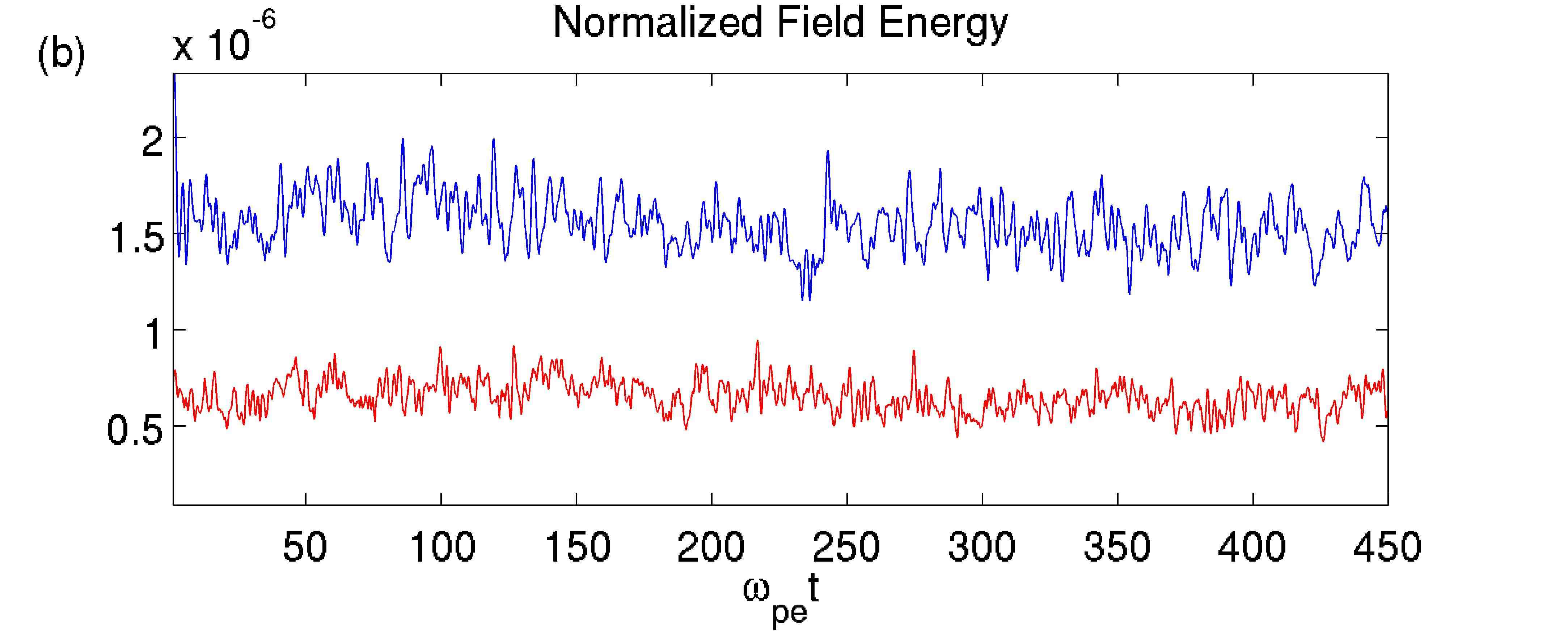}
\caption{Effect of the Multi Level Multi Domain system on the total field energy in the simulation of a Maxwellian plasma. Total field energy normalized to the total domain (fields plus particles) energy at $\omega_{pe}t=0$ for the (a) Multi Level Multi Domain and (b) reference case respectively. The blue line is relative to the coarser grid, the red line to the refined grid.}
\label{fig:MAX_Energy}
\end{figure}

To test how grid coupling influences the simulation evolution in absence of notable plasma activity, a Maxwellian plasma is simulated across the two domains. Electrons and ions with a realistic mass ratio of  $m_{i}/m_{e}=1836$, a thermal velocity of $v_{th}/c=0.2$ in all directions ($c$ is the speed of light, used as normalization factor for the velocities) and no drift are immersed in a domain of size $L_{x, g_{l0}}/d_{e}=84$, with $d_{e}$ the electron skin depth. A refined grid with Refinement Factor RF=4 overlaps the coarser grid at $21 \leq x/d_{e} \leq 42$. The information exchange between the two grid is done accordingly to Sec.~\ref{sec:InterDomainInter}, with a PRA (see Sec.~\ref{sec:particleRepop}) extending for one cell into the active part of the refined grid. Both grids have $280$ cells, the time step is $\omega_{pe}dt=0.15$, with $\omega_{pe}$ the electron plasma frequency, and the simulation is carried on for $3000$ cycles. A reference case with two levels simulated independently, both with periodic conditions for fields and particles, is also shown to help assessing the impact of the MLMD algorithm on this basic simulation.\\
Fig.~\ref{fig:MAX_Energy} depicts the total field energy normalized to the initial energy (hence the low value: the energy of the system is mostly stored in particles) for the MLMD (panel a) and the reference (panel b) case respectively, with the blue line referred to the coarser and the red line to the refined grid in both panels.

\begin{figure}[ht]
\centering
 \includegraphics[width=8cm]{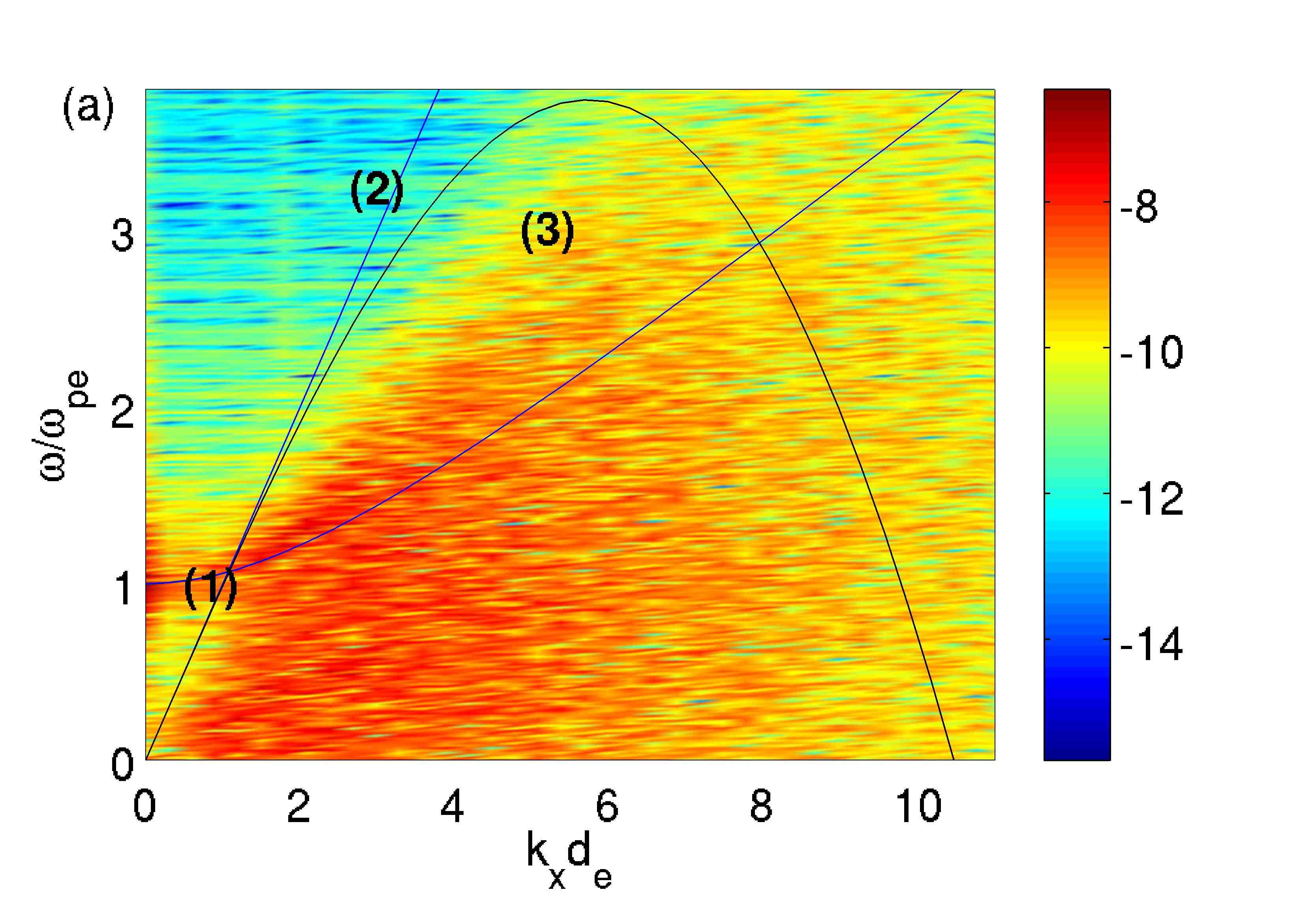}
 \includegraphics[width=8cm]{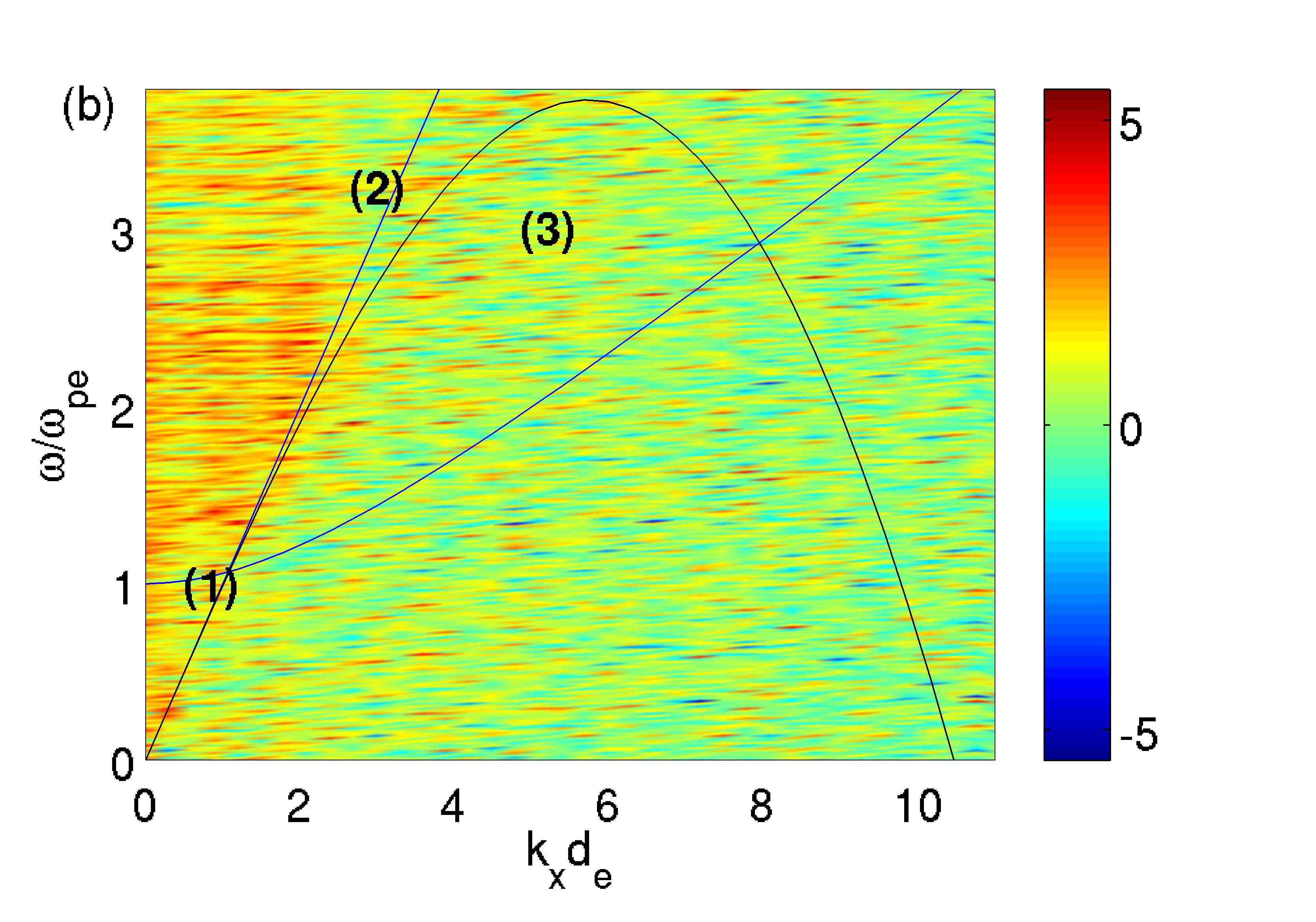}
\caption{Effect of the Multi Level Multi Domain system on the spectrum of the longitudinal electric field in the simulation of a Maxwellian plasma. (a) Fast Fourier Transform of $E_{x, N,g_{l1}}$ in the Multi Level Multi Domain simulation and (b) difference with the corresponding spectrum in the reference case.}
\label{fig:MAX_Spectrum}
\end{figure}

Notice that in the reference case, where no interlevel communications are present (panel b), the field energy stabilizes around different values for the coarser and refined grid, with the coarser grid exhibiting higher energy values. Instead, in the MLMD case (panel a) the two grids tend to converge to the coarser grid energy level after a transient time which corresponds to the time needed for the repopulated particles to overcome in number the native ones which exit the grid by thermal motion. The fact that the refined level energy raising to the coarser grid values is connected to particle repopulation is confirmed also by check simulations with MLMD conditions on the refined grid fields but periodic conditions on the refined grid particles, which exhibit energy levels comparable to the reference case.\\This evolution of the field energy in the MLMD system is considered a symptom of the efficient coupling between the refined and coarser grid, which is mostly achieved through particle repopulation. A spectral analysis of $E_{x, N,g_{l1}}$ and $E_{x, P,g_{l0}}$ in the MLMD system, additionally, shows that this effect is not detrimental to the physical significance of the simulation.\\
Fig.~\ref{fig:MAX_Spectrum}, panel a, shows the Fast Fourier Transform of $E_{x, N,g_{l1}}$ in the MLMD system, while Fig.~\ref{fig:MAX_Spectrum}, panel b, depicts the difference of it with the corresponding spectrum for the reference case.

Notice that Fig.~\ref{fig:MAX_Spectrum}a exhibits all the features expected from an implicit PIC simulation of a Maxwellian plasma: the Langmuir wave (marked as (1)) and the dispersion relation for the light wave (marked as (2)) as reproduced by an implicit PIC code\citep{brackbill82}(marked as (3)) are both visible and not modified in the MLMD case when compared to the reference one, as shown in Fig.~\ref{fig:MAX_Spectrum}b. The extra energy in the refined grid of the MLMD case is stored at low $k/d_{e}$s in an area unaffected by relevant features and thus proves not to damage the physical relevance of the simulation.\\
A similar analysis carried out for the coarser level shows that no relevant difference is observable between the spectrum of $E_{x, P,g_{l0}}$ in the MLMD and reference case.

\subsection{Testing the development of instabilities and the continuity of stable particle structures across the grid boundary: Weibel instability simulation}
\label{sec:weibel}

\begin{figure}[ht]
\centering
 \includegraphics[width=8cm]{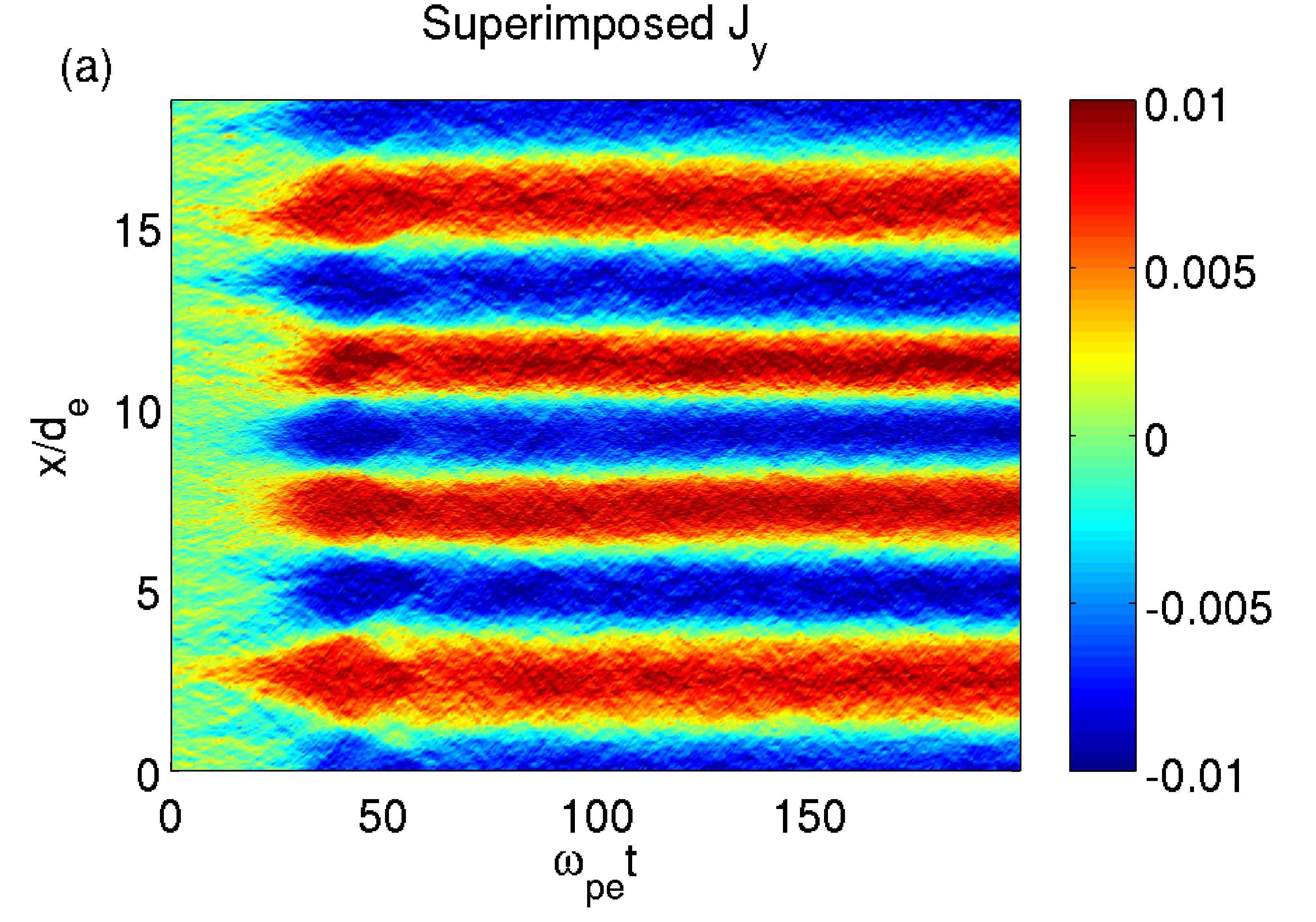}
 \includegraphics[width=8cm]{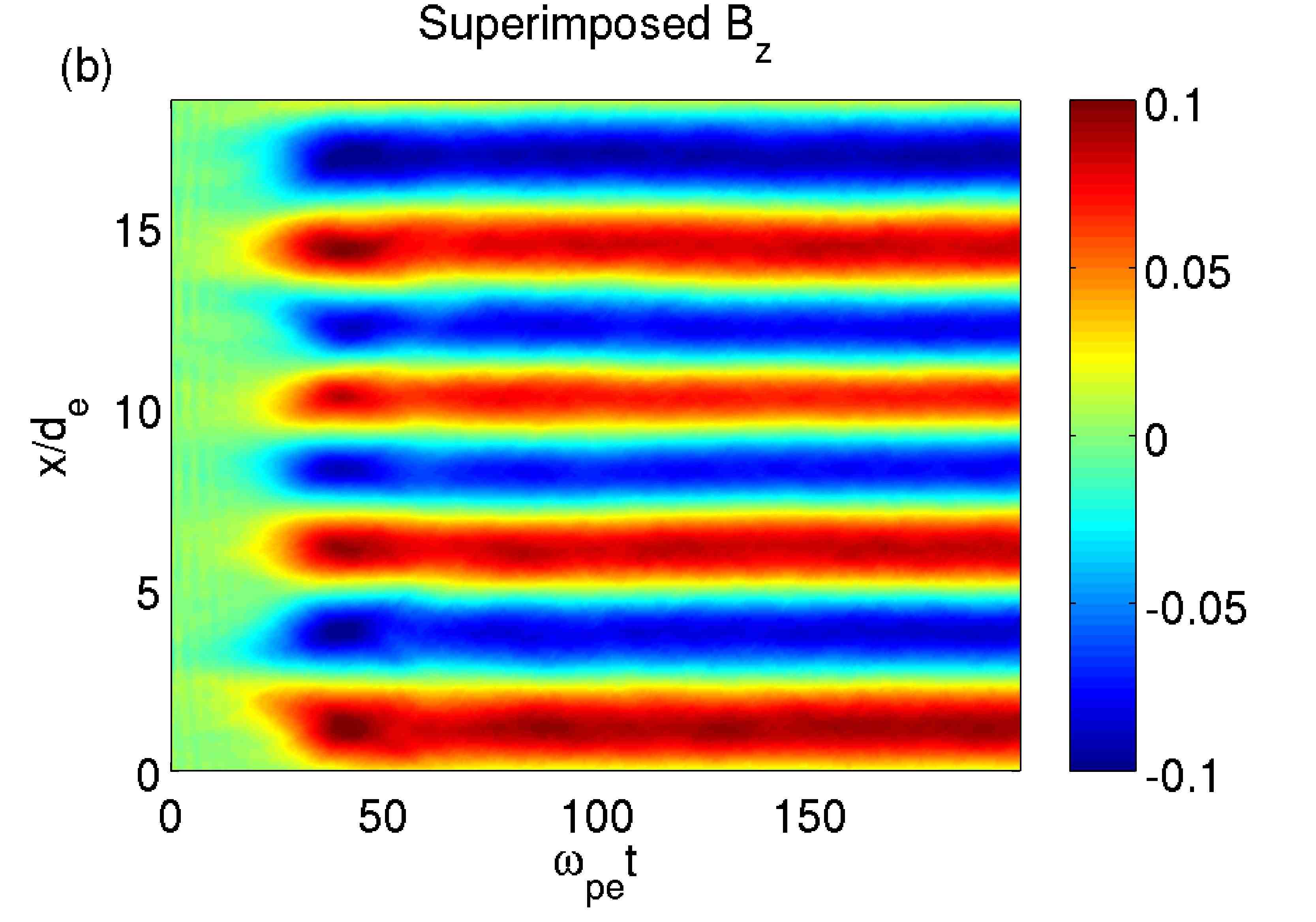}
\caption{Relevant physical quantities in the Multi Level Multi Domain simulation of the Weibel instability. (a) Current $J_{y,g_{l0}}$ with superimposed $J_{y,g_{l1}}$ and (b) magnetic field $B_{z,P,g_{l0}}$ with superimposed $B_{z,N,g_{l1}}$ at $6.13 \leq x/d_{e} \leq 10.75$. }
\label{fig:weibel}
\end{figure}

The Weibel instability is excited by an anisotropy in the thermal velocity of particles, i.e. by the presence of a preferential, warmer direction in the particle distribution function. The resulting particle motion produces a current in the direction of higher temperature which in turn grows, through the Ampere's equation, a perpendicular magnetic field. The instability eventually saturates due to magnetic trapping processes which lead to the formation of characteristic structures in the phase spaces of electrons, with trapping points being formed at the zeros of the magnetic field and a consequent zig-zag distribution emerging in the $v_{y}/c$ $vs$ $x/d_{e}$ electron phase space\citep{weibel59, stockem09, innocenti11}. In a 1D system simulated in the $x$ direction with $y$ the direction of higher particle temperature, a current $J_{y}$ is formed in the warmer $y$ direction and a magnetic field $B_{z}$ is consequently grown.\\
It is investigated here how the magnetic field is originated in the MLMD system and the continuity of the particle structures across the grid boundaries.\\
The Weibel instability is excited by setting the thermal velocities for electrons to $v_{th,y}/c=0.2$ and $v_{th,x}/c=v_{th,z}/c=0.05$ in the $y$ and $x$ and $z$ direction respectively. The ions, with a mass ratio of $m_{i}/m_{e}=1836$, are loaded with the same temperature and temperature anisotropy as the electrons. The time step is $\omega_{pe}dt=0.1$, the coarser grid has length $L_{x, g_{l0}}/d_{e}=18.47$ and both the coarser and the refined grids have $280$ cells. The refined grid overlaps the coarser grid at $6.13 \leq x/d_{e} \leq 10.75$, the Refinement Factor is $RF=4$ and three PRA cells are used in the active domain of the refined grid. Remember that the number of PRA cells to be used is decided a priori according to the spatial and temporal resolution of the refined grid and the expected maximum velocity of particles. More sophisticated algorithms will be devised for future implementations.\\
From the linear theory, the dominant wavenumber in such a configuration is $kd_{e}= 1.36$ with growth rate $\gamma / \omega_{pe}=0.123$. In the presented simulation geometry, the dominant mode $n$, with $L_{x, g_{l0}}=n\frac{2\pi}{k}$, is $n=4$. \\
Fig.~\ref{fig:weibel} shows the current $J_{y}$ (panel a) and the magnetic field $B_{z}$ (panel b) in the aforementioned MLMD case, with the refined grid fields superimposed to the coarser grid ones. Notice that both $J_{y}$ and $B_{z}$ develop as expected~\citep{stockem09}, with a calculated growth rate for $kd_{e}=1.36$ equal to $\gamma / \omega_{pe}= 0.123$, the theoretical value, and that no discontinuity is noticeable at the boundaries of the refined grid. It is therefore possible to state that the correct development of the instability is preserved by the MLMD algorithm.

\begin{figure}[ht]
\centering
\includegraphics[width=9cm]{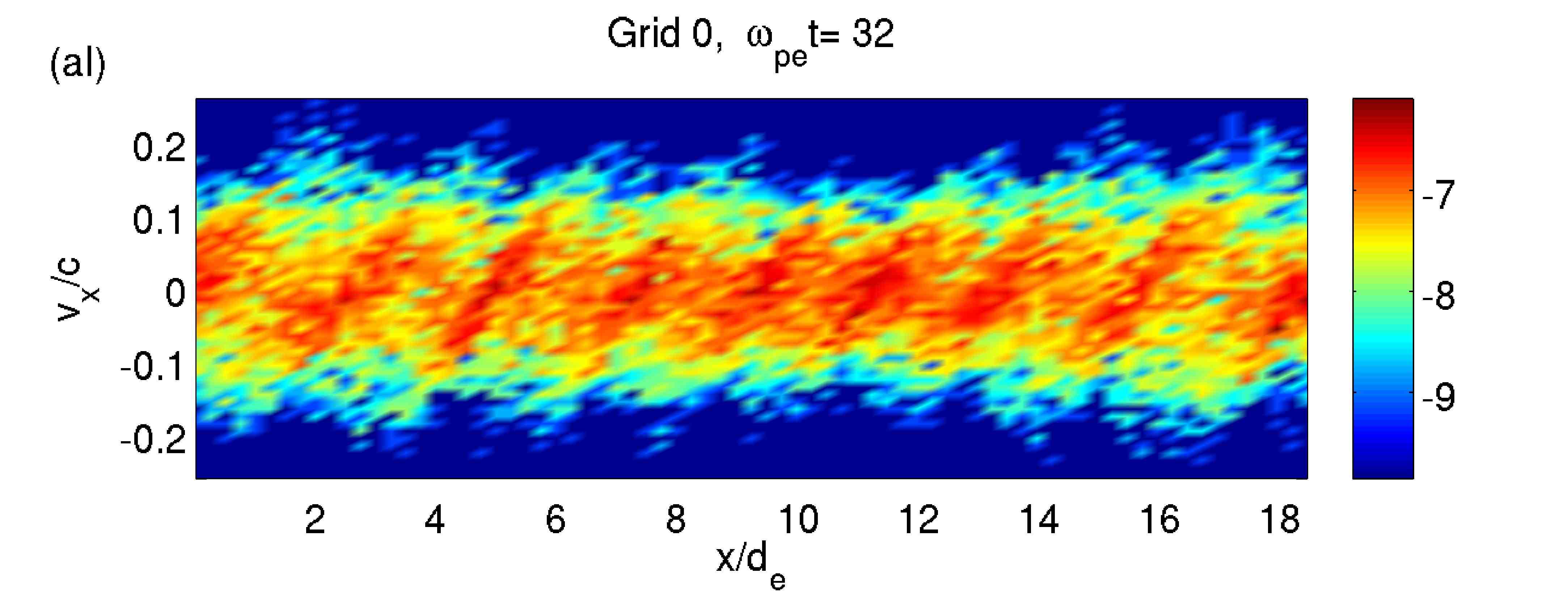}
\includegraphics[width=9cm]{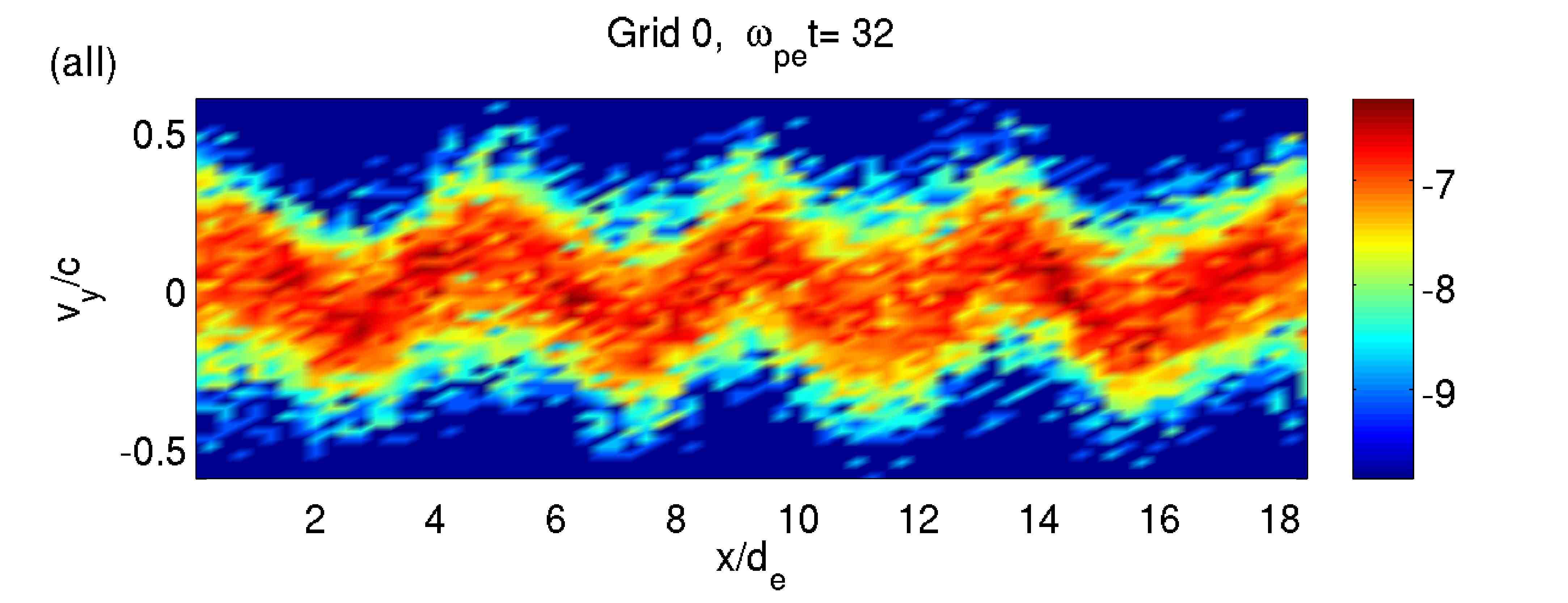}
\includegraphics[width=9cm]{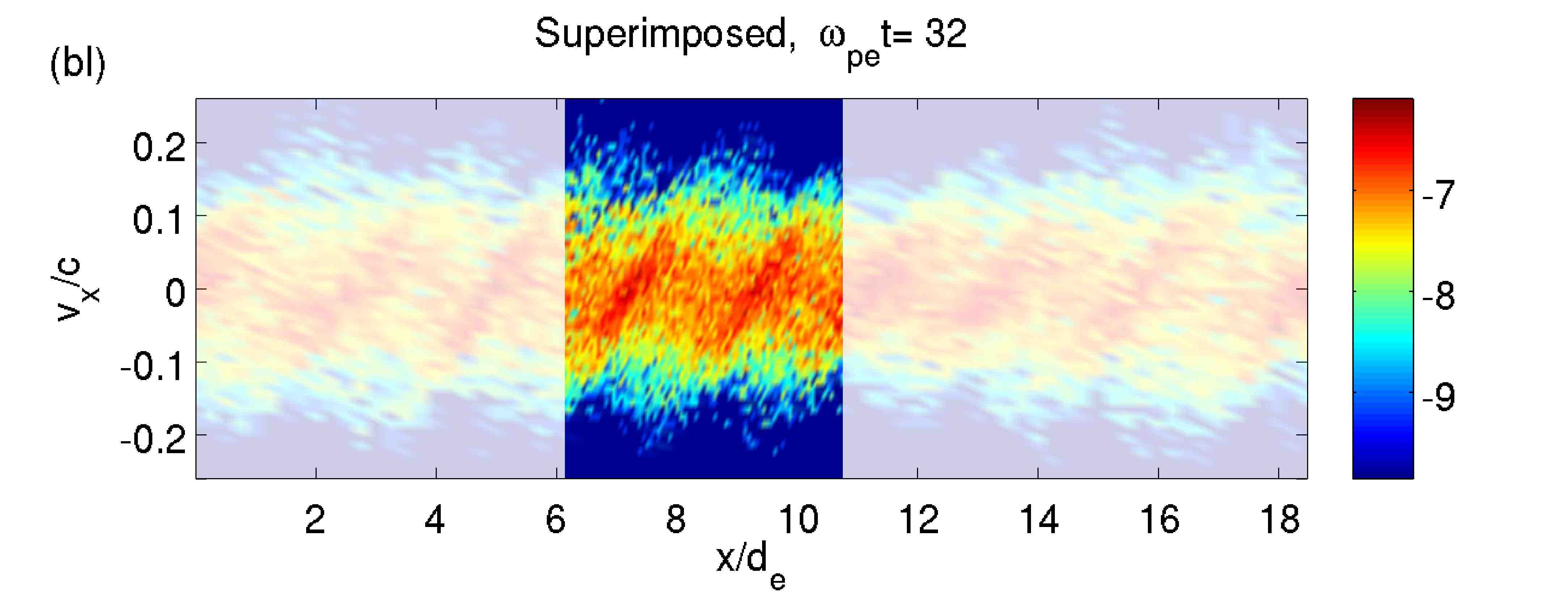}
\includegraphics[width=9cm]{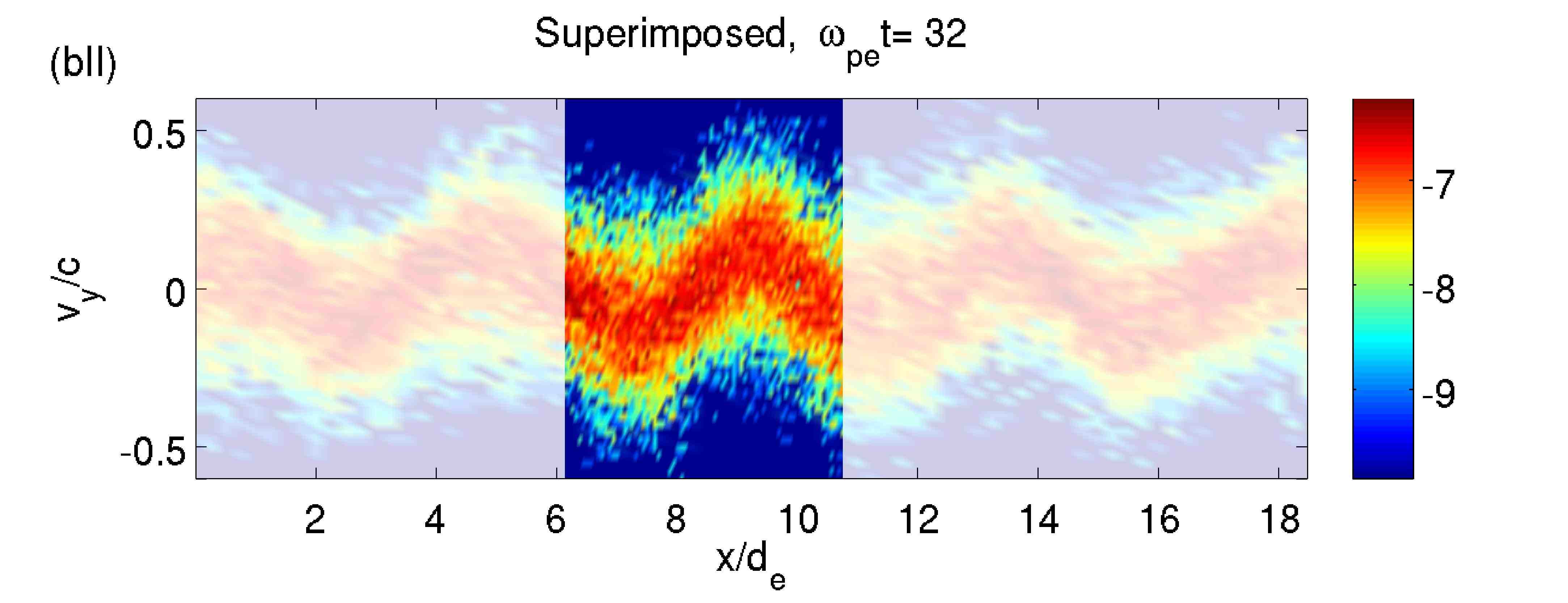}
\caption{Phase spaces before saturation for the Multi Level Multi Domain simulation of the Weibel instability. (a) Phase space $v_{x}/c$ $vs$ $x/d_{e}$ (aI) and $v_{y}/c$ $vs$ $x/d_{e}$ (aII) for the coarser grid and (b) phase space $v_{x}/c$ $vs$ $x/d_{e}$ (bI) and $v_{y}/c$ $vs$ $x/d_{e}$ (bII) for the refined grid superimposed to the coarser grid phase spaces, shown in transparency, at $\omega_{pe}t=32$.}
\label{fig:pt_bsat}
\end{figure}

\begin{figure}[ht]
\centering
\includegraphics[width=9cm]{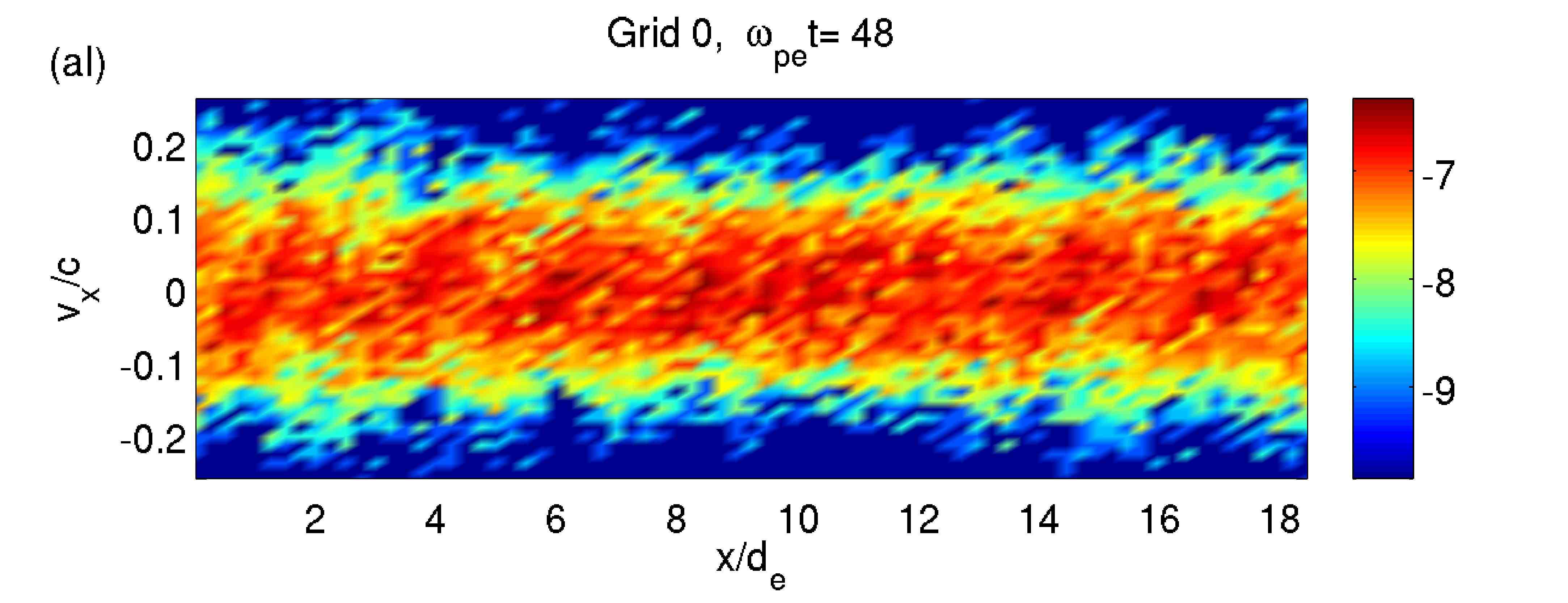}
\includegraphics[width=9cm]{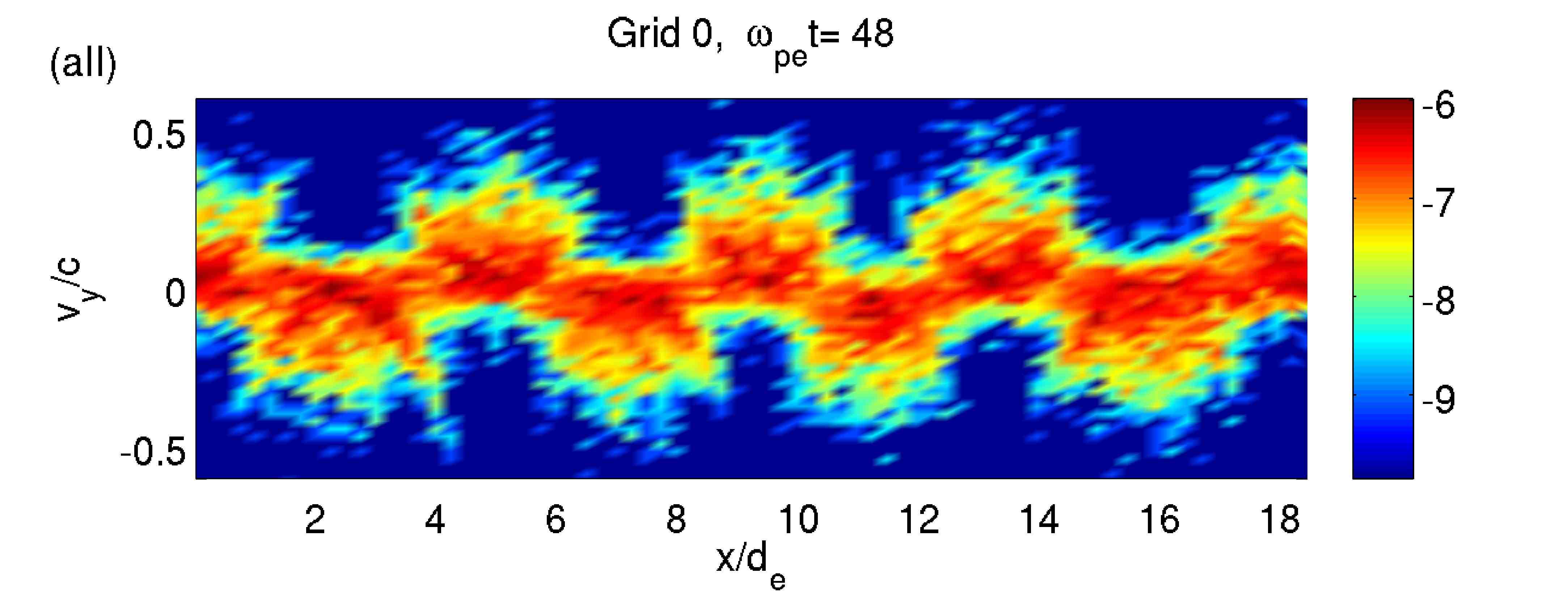}
\includegraphics[width=9cm]{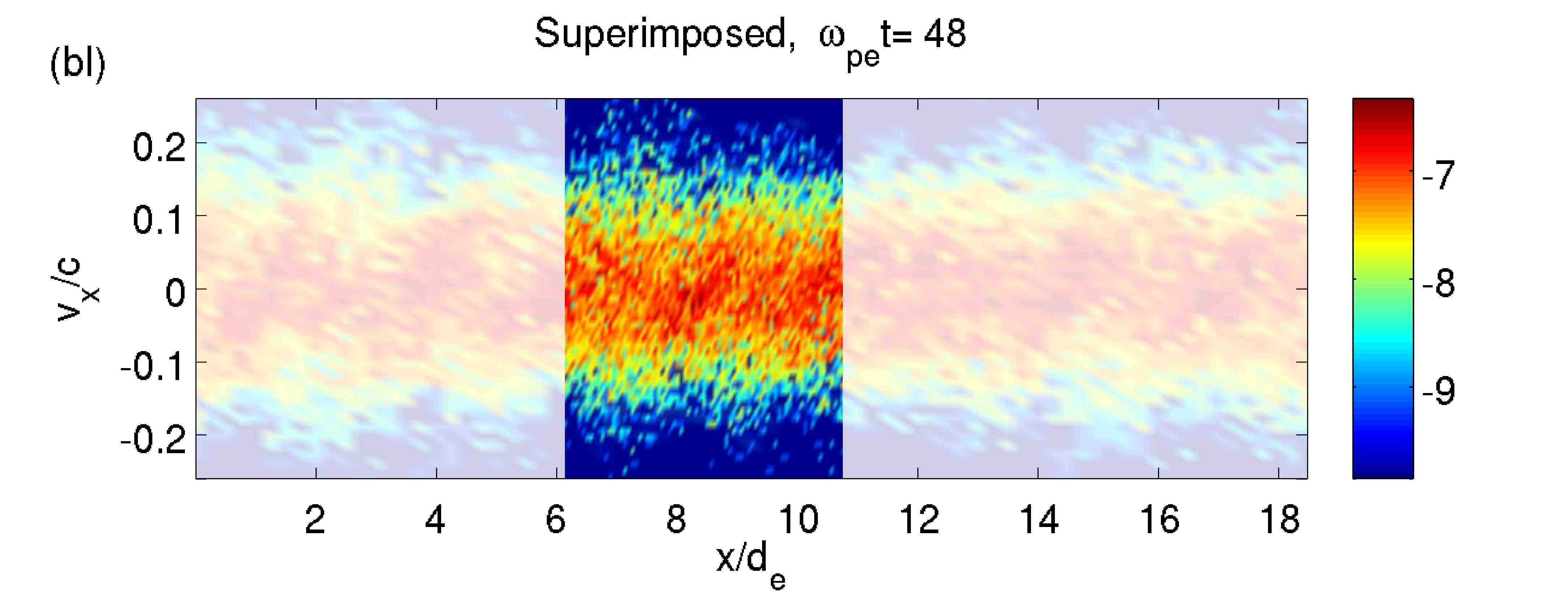}
\includegraphics[width=9cm]{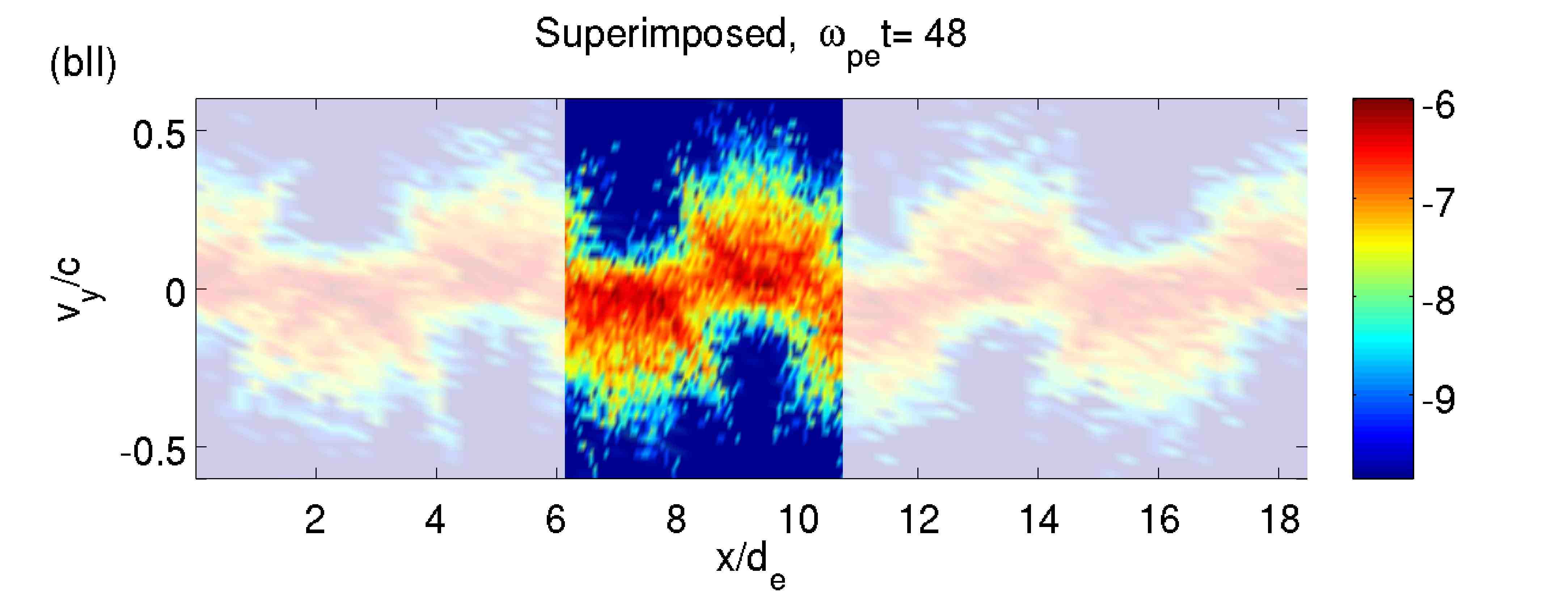}
\caption{Phase spaces after saturation for the Multi Level Multi Domain simulation of the Weibel instability. (a) Phase space $v_{x}/c$ $vs$ $x/d_{e}$ (aI) and $v_{y}/c$ $vs$ $x/d_{e}$ (aII) for the coarser grid and (b) phase space $v_{x}/c$ $vs$ $x/d_{e}$ (bI) and $v_{y}/c$ $vs$ $x/d_{e}$ (bII) for the refined grid superimposed to the coarser grid phase spaces, shown in transparency, at $\omega_{pe}t=48$.}
\label{fig:pt_aftersat}
\end{figure}

More interesting is however the inspection of the phase space plots, with the aim of checking if the electron trapping structures are correctly developed and how they are dealt with across the grid boundaries. \\
Fig.~\ref{fig:pt_bsat} and~\ref{fig:pt_aftersat} show, in panel aI and aII respectively, the phase space $v_{x}/c$ vs $x/d_{e}$ and $v_{y}/c$ vs $x/d_{e}$ for the electrons of the coarser grid alone, while in panel bI and bII the refined grid phase spaces are superimposed to the coarser grid ones to check for particle structure continuity. Fig.~\ref{fig:pt_bsat} refers to time $\omega_{pe}t=32$, when particle trapping starts becoming evident, while Fig.~\ref{fig:pt_aftersat} is taken after the saturation of the mode with $n=4$, at $\omega_{pe}t=48$ (saturation occurs at $\omega_{pe}t=41$).\\ 
Note that the particle structures develop as expected from previous studies\citep{stockem09, innocenti11} and as in the reference case, not shown here, with electron trapping points being formed at the zeros of $B_{z}$ in the simulated $x$ direction and a zig zag distribution emerging in $v_{y}/c$ $vs$ $x/d_{e}$ in accordance with the current structure. Notice that the particle structures are coherent between the refined and the coarser level and continuous among the refined and coarser grid boundaries. The $v_{y}/c$ vs $x/d_{e}$ phase space plot in Fig.~\ref{fig:pt_aftersat} panel bII reveals a particularly impressive grid coupling: even the smallest scale electron features are correctly captured across the grid boundaries.

\subsection{Testing the continuity of moving particle structures across the grid boundary: two stream instability simulation}
\label{sec:2streams}

\begin{figure}[ht]
\centering
 \includegraphics[width=8cm]{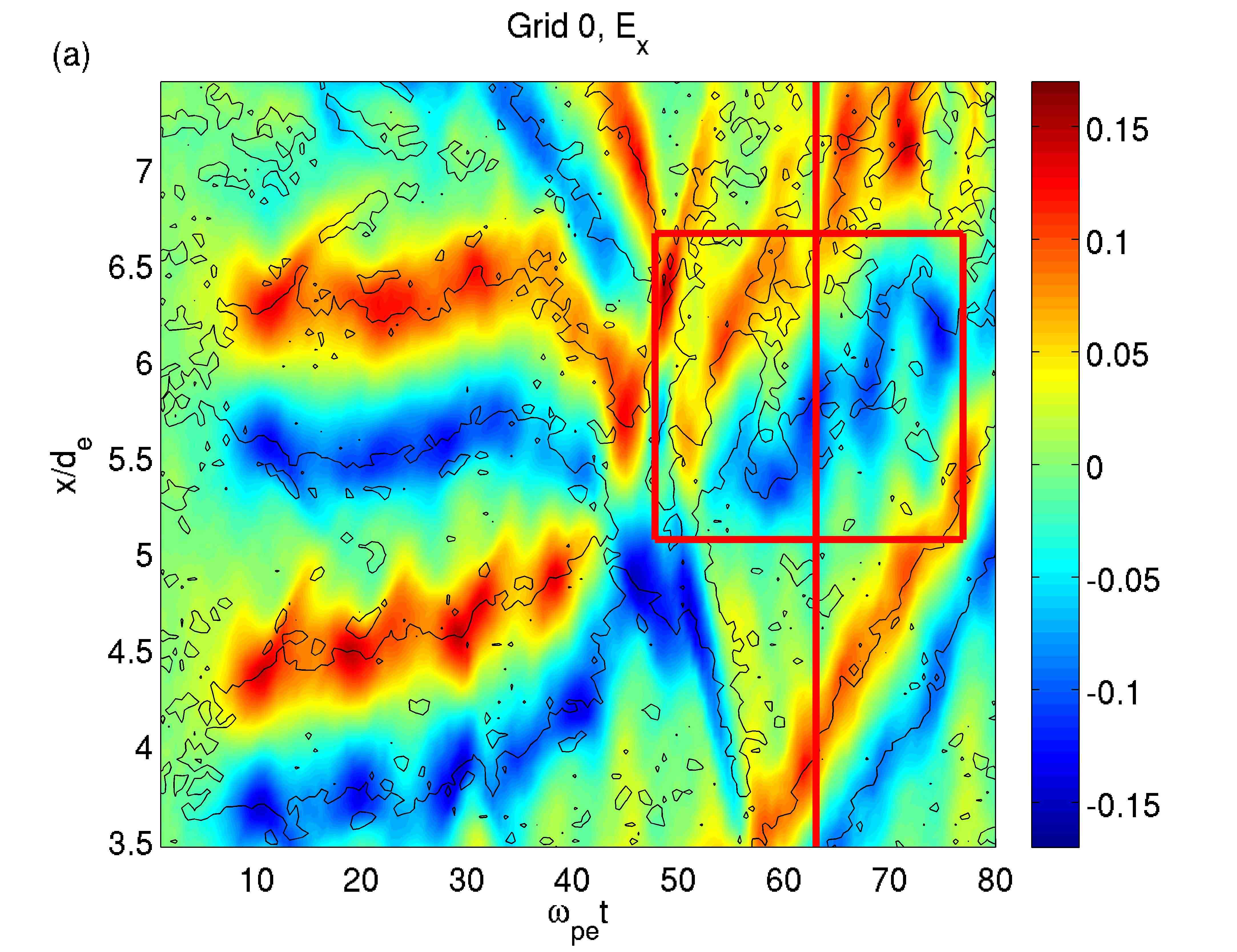}
 \includegraphics[width=8cm]{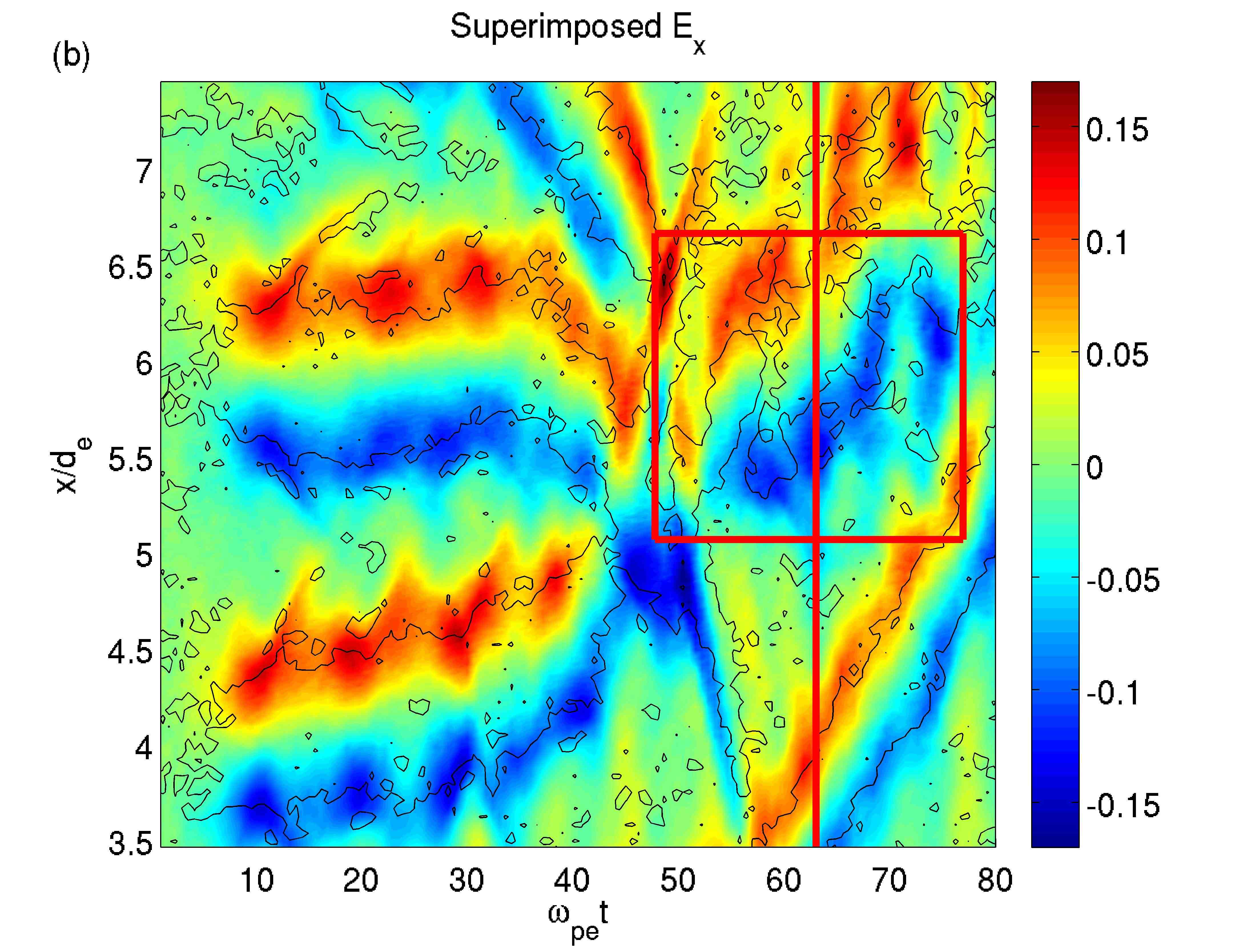}
\caption{Electric field mismatch between the levels in the overlapping area for the Multi Level Multi Domain simulation of the two stream instability. (a) Electric field $E_{x,P,g_{l0}}$ on the coarser grid alone and (b) with superimposed the refined grid field $E_{x,N,g_{l1}}$ at $3.96-0.5 \leq x/d_{e} \leq 6.95+0.5$, with $x/d_{e}=3.96$ and $x/d_{e}=6.95$ being the refined grid boundaries. The contour of the coarser level density $n_{g_{l0}}$ is superposed to both panels. The red rectangle and, in particular, the red line at $\omega_{pe}t=62$ mark an area of rather pronounced differences between $E_{x,P,g_{l0}}$ and $E_{x,N,g_{l1}}$.}
\label{fig:Ex2streamANDdensities}
\end{figure}

While in the case of the Weibel instability the continuity across the refined and coarser grid boundaries of the electron structures in phase space could be argued to be a consequence of the fact that the trapping positions for particles are located at fixed points, similar arguments cannot be made in the case of a MLMD simulation of a two stream instability\citep{chen}. In this case in fact the evolution and merging of the generated electron holes cannot be determined a priori and the structures move through the domain, since the instability has a real frequency.\\
In the case here presented, the two stream instability is excited through an electron stream velocity of $V_{x}/c=\pm0.2$ in the simulated direction, while the thermal velocities for ions, with realistic mass ratio, and electrons are $v_{th}/c=0.04$. The coarser grid has length $L_{x, g_{l0}}/d_{e}=11.94$, the time step is $\omega_{pe} dt=0.1$ and $2000$ computational cycles are executed. The refined grid has Refinement Factor $RF=4$, extends at $3.96 \leq x/d_{e} \leq 6.95$ and nine PRA cells are used (such an high number of PRA cells is due to the high velocities expected in the system after the development of the instability). 
Fig.~\ref{fig:Ex2streamANDdensities} depicts the electric field $E_{x, P, g_{l0}}$ for the coarser grid in panel a, while in panel b the refined grid $E_{x, N, g_{l1}}$ is superimposed at $3.96 \leq x/d_{e} \leq 6.95$. In both cases, the contour of the coarser field density $n_{g_{l0}}$ is shown on top of the electric field. This picture, contrarily to before, does not depict the entire coarse grid domain and simulation duration, but just a fraction of both: $3.96-0.5 \leq x/d_{e} \leq 6.95+0.5$, with a $\pm0.5/d_{e}$ distance from the finer grid boundary, and $\omega_{pe}t \leq 80$, when the electron holes merging activity is more intense. The aim is to show that while, as usual, field continuity is granted at the boundary between the coarser and refined grid, a certain mismatch in $E_{x}$ is visible \textsl{inside} the overlap area.
Focus, for example, on the area highlighted in red in both panels: notice that the traces of negative electric field, even if roughly similar, exhibit small differences.

\begin{figure}[ht]
\centering
 \includegraphics[width=9cm]{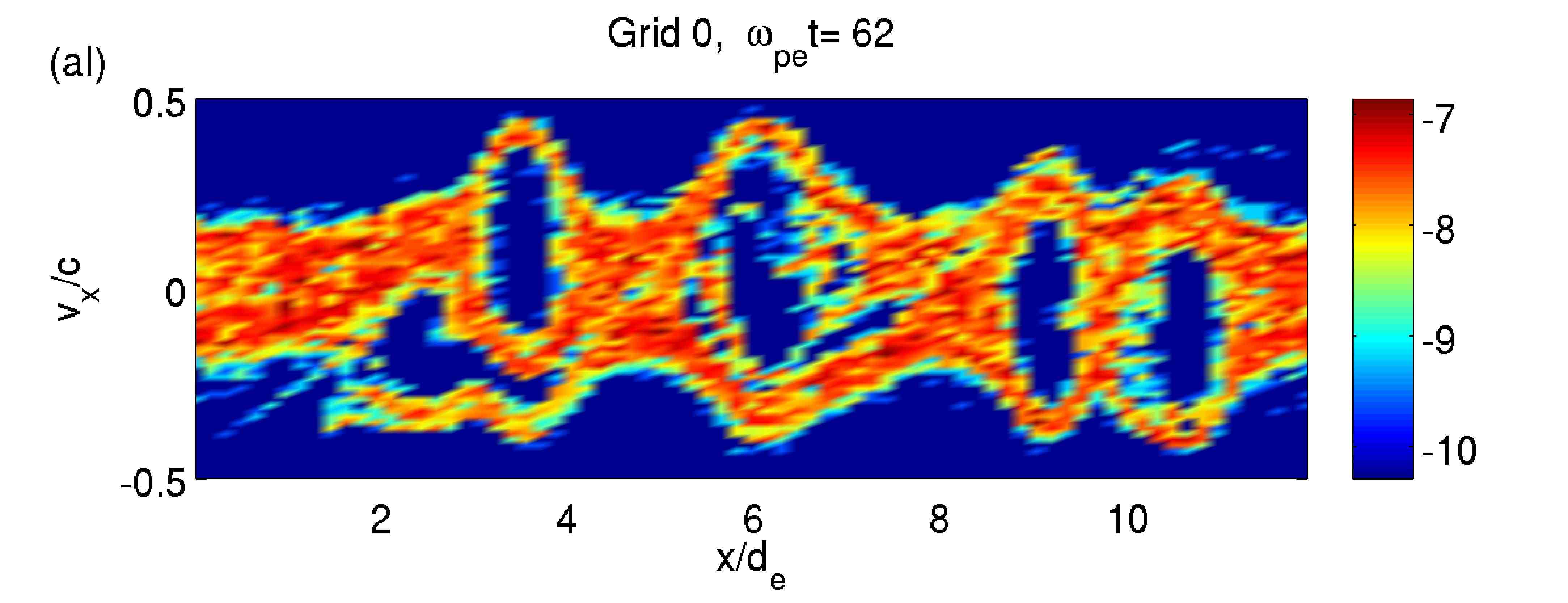}
 \includegraphics[width=9cm]{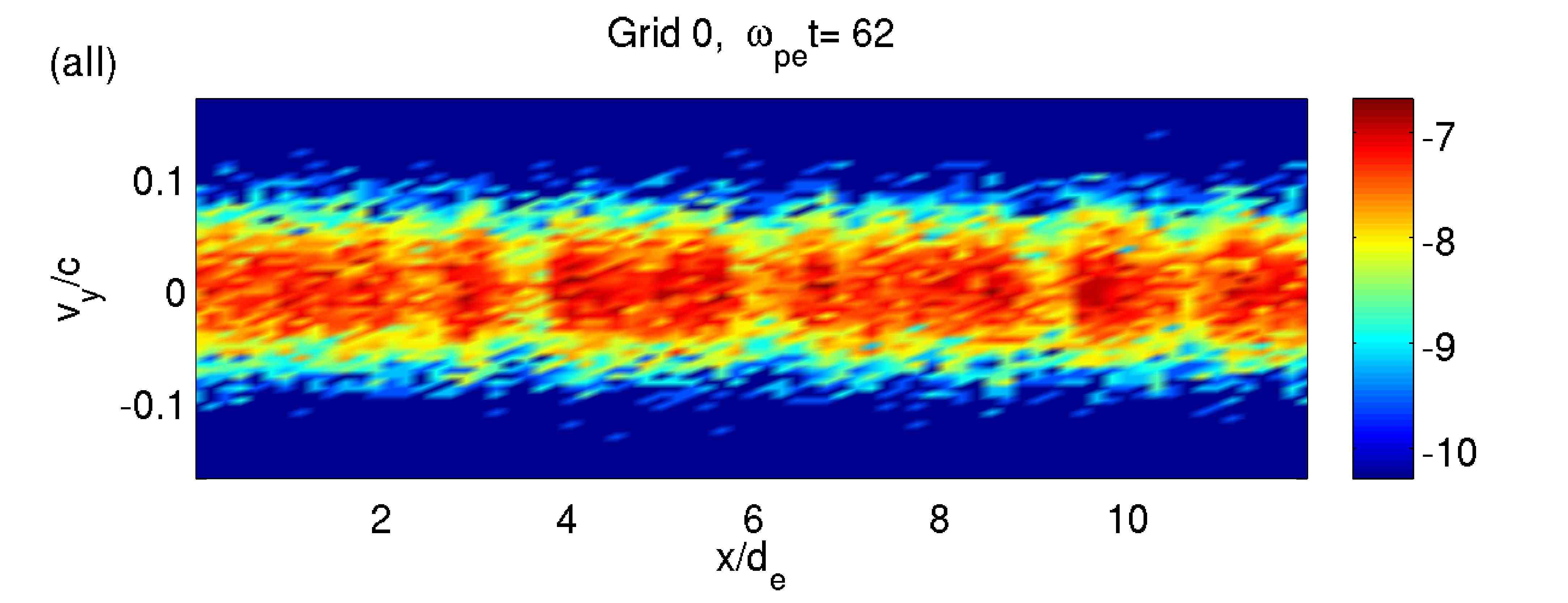}
 \includegraphics[width=9cm]{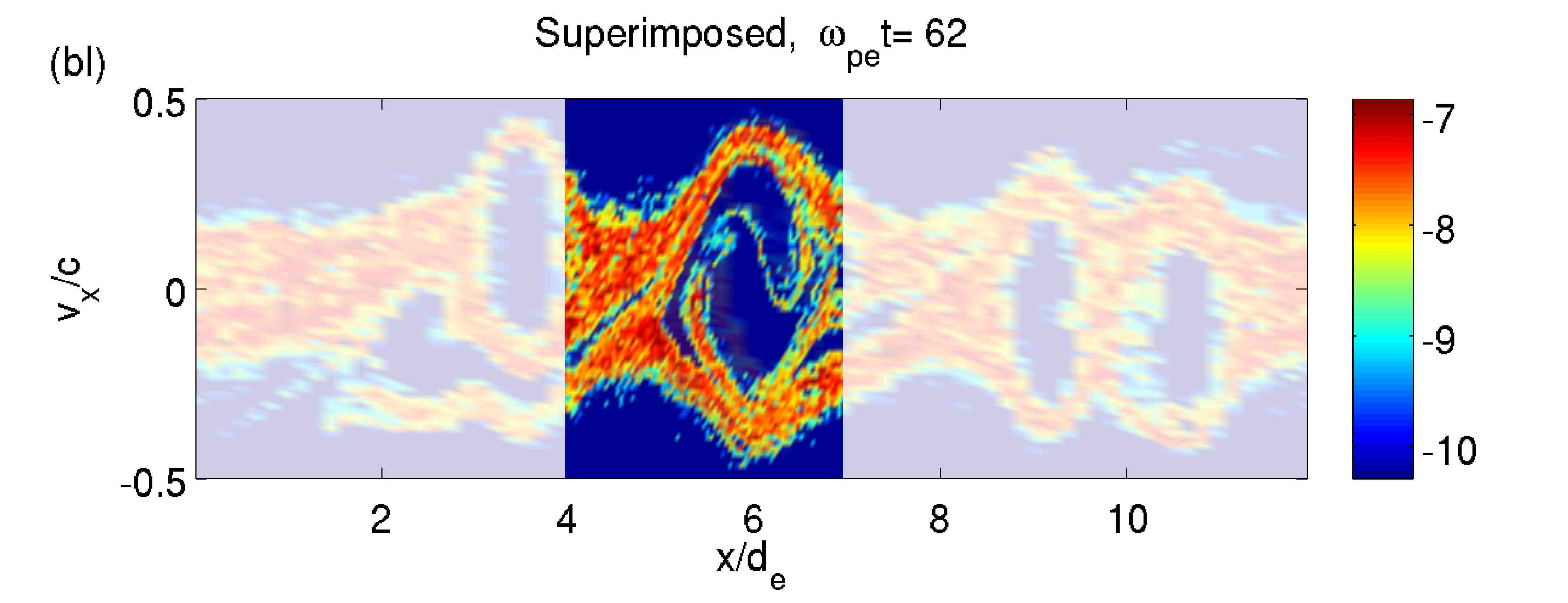}
 \includegraphics[width=9cm]{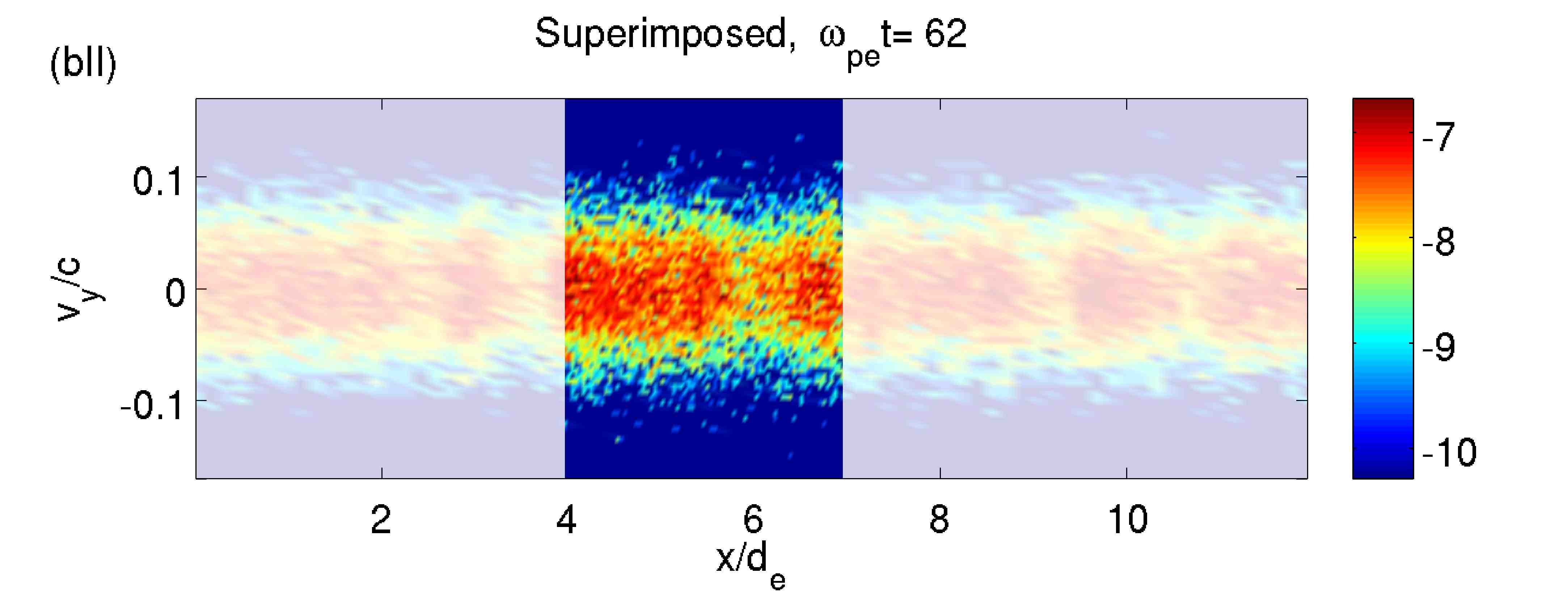}
 
\caption{Phase spaces at a time of rather pronounced differences between the longitudinal electric fields in the different levels for the Multi Level Multi Domain simulation of the two stream instability. (a) Phase space $v_{x}/c$ $vs$ $x/d_{e}$ (aI) and $v_{y}/c$ $vs$ $x/d_{e}$ (aII) for the coarser grid and (b) phase space $v_{x}/c$ $vs$ $x/d_{e}$ (bI) and $v_{y}/c$ $vs$ $x/d_{e}$ (bII) for the refined grid superimposed to the coarser grid phase spaces, shown in transparency, at $\omega_{pe}t=62$.}
\label{fig:2stream_ps_1}
\end{figure}

\begin{figure}[ht]
\centering
 \includegraphics[width=9cm]{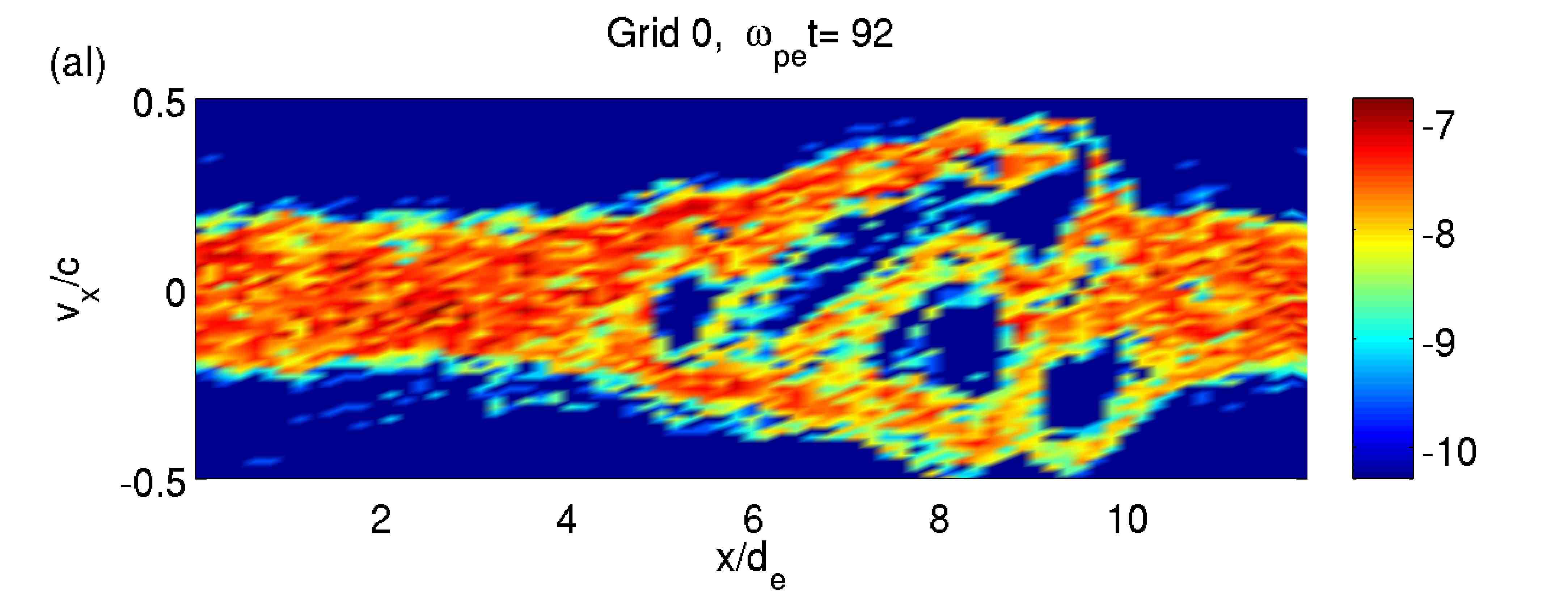}
 \includegraphics[width=9cm]{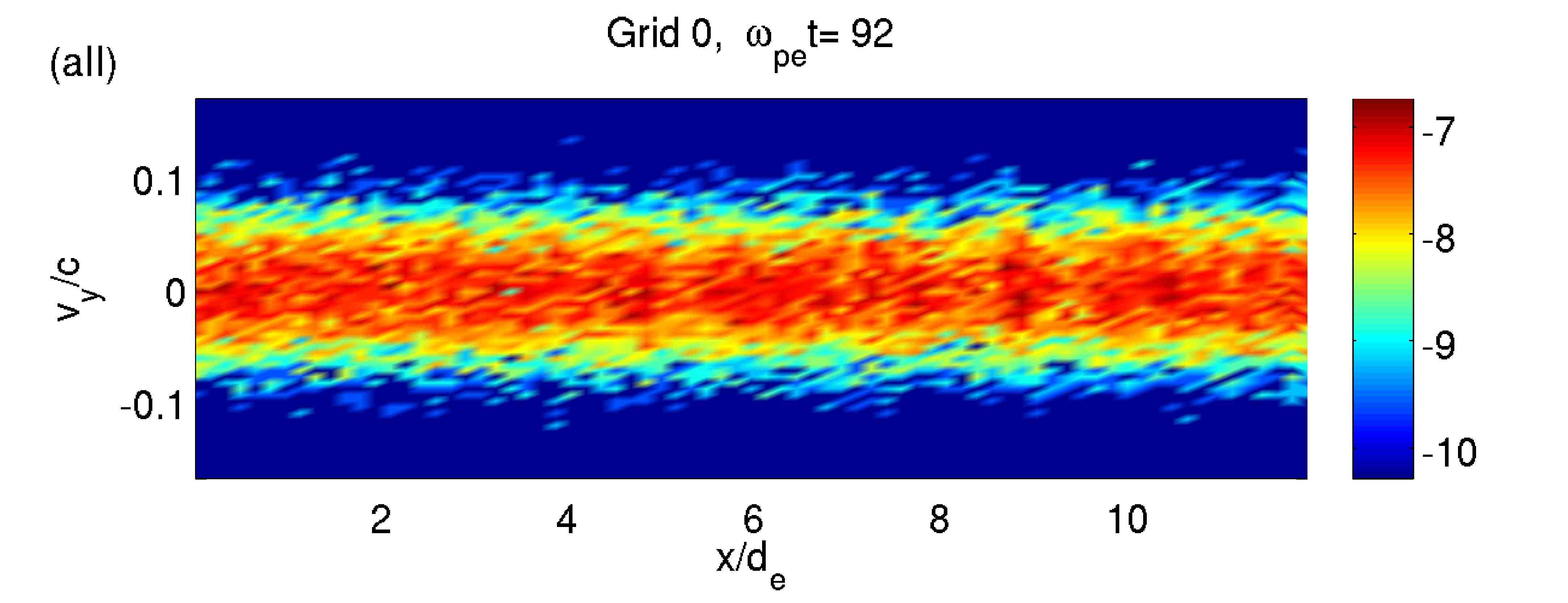}
 \includegraphics[width=9cm]{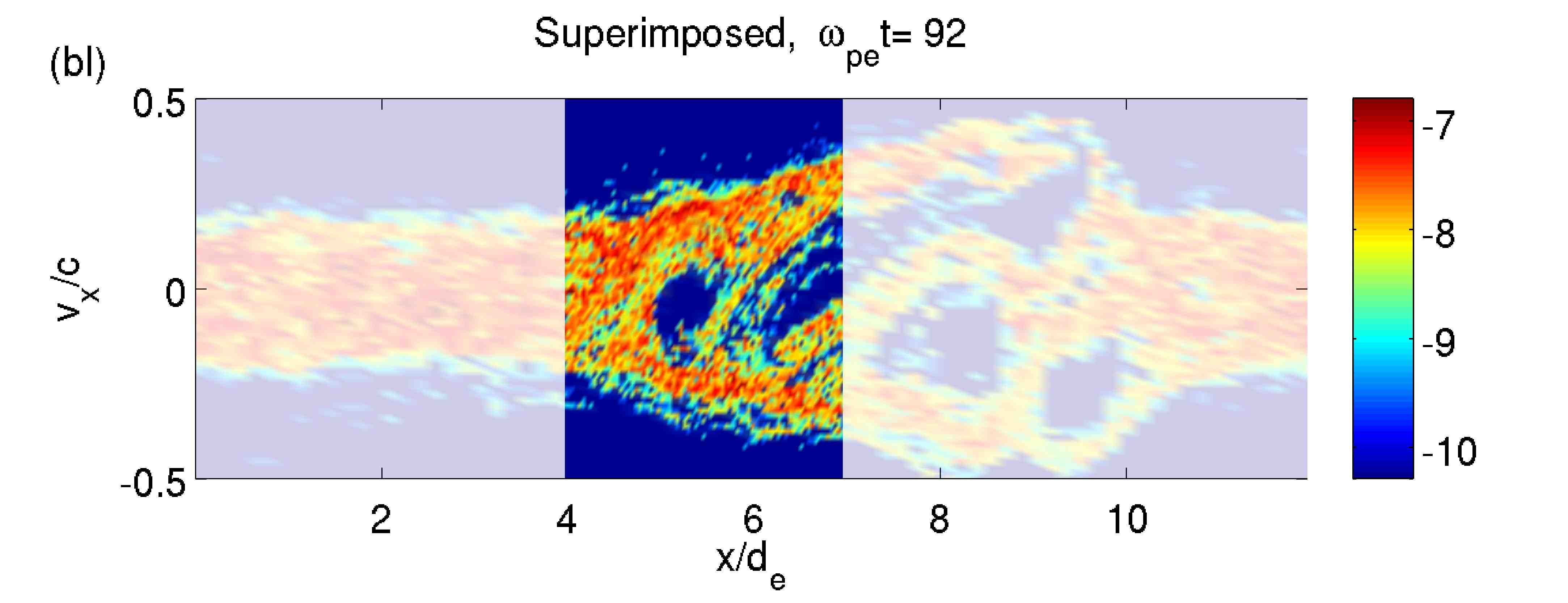}
 \includegraphics[width=9cm]{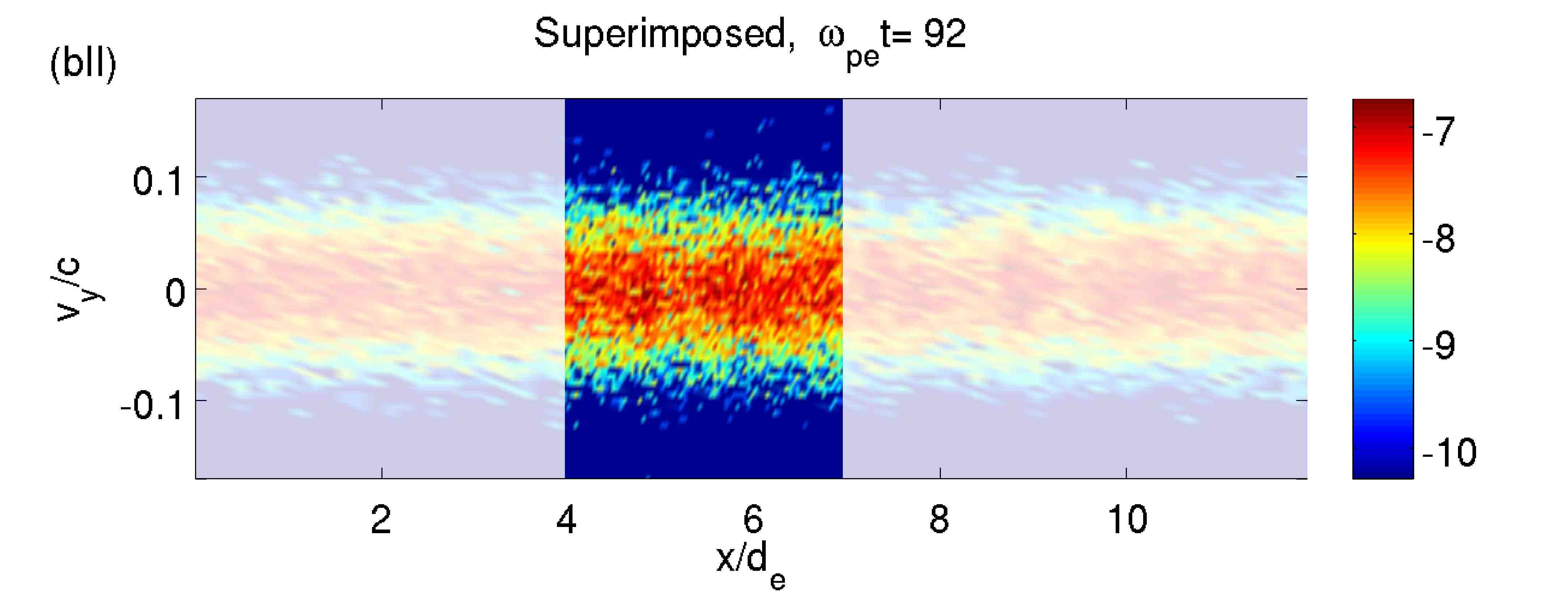}
\caption{Phase spaces at a time of minimal differences between the longitudinal electric fields in the different levels for the Multi Level Multi Domain simulation of the two stream instability. (a) Phase space $v_{x}/c$ $vs$ $x/d_{e}$ (aI) and $v_{y}/c$ $vs$ $x/d_{e}$ (aII) for the coarser grid and (b) phase space $v_{x}/c$ $vs$ $x/d_{e}$ (bI) and $v_{y}/c$ $vs$ $x/d_{e}$ (bII) for the refined grid superimposed to the coarser grid phase spaces, shown in transparency, at $\omega_{pe}t=92$.}
\label{fig:2stream_ps_2}
\end{figure}

This mismatch is due to the fact that the electron holes that produce the traces in $E_{x}$ evolve slightly differently in the two grids, since the coarser grid forcing over the refined grid is limited to the boundaries for fields and to the few PRA cells for particles. Therefore, some differences in the particle evolution and, consequently, in the traces in $E_{x}$ in the overlap area far away from the boundaries have to be expected.
The above mentioned mismatch between the electric field evolution at different levels is removed if the electric field is projected by substitution instead of averaging.
However, as already argumented in Sec.~\ref{sec:upwards}, projecting the electric field by substitution proves detrimental in cases where the major simulated dynamics are not electrostatic. For this reason, for sake of generality of implementation, the projection of the electric field is done by average as described in Sec.~\ref{sec:upwards}, with the caveat that particle moments and fields are consistent within the grids (i.e., $E_{P,g_{l}}$ is consistent with the density $n_{g_{l}}$ on the same level) and not across the grids (i.e., $E_{N,g_{l+1}}$ is not necessarily consistent with the density $n_{g_{l}}$ in the overlap area of the coarser grid). 
This is well visible in Fig.~\ref{fig:Ex2streamANDdensities}, where the coarser grid density $n_{g_{l0}}$ is superimposed to the coarser grid field in panel a and to the refined grid field in panel b. It is well known that the peak density for electron holes is registered at the center of the electric field bipolar trace\citep{eliasson05}: see, focusing for example on the area enlightened by the red rectangle, that $n_{g_{l0}}$ is consistent with the coarser field $E_{x,P,g_{l0}}$, not with the refined grid one, $E_{x,N,g_{l+1}}$, even if the differences are really minimal to catch. $n_{g_{l1}}$, of course, is consistent with the refined grid fields.\\
Let us check now what this means in term of phase space plots.\\
Fig.~\ref{fig:2stream_ps_1} and Fig.~\ref{fig:2stream_ps_2} depict the phase space plots $v_{x}/c$ $vs$ $x/d_{e}$ (marked as I) and $v_{y}/c$ $vs$ $x/d_{e}$ (marked as II) for the above mentioned simulation at two different times, $\omega_{pe}t=62$ and $\omega_{pe}t=92$. In both cases, panel a refers to the phase spaces of the coarser grid, while in panel b the phase spaces for the refined grid are superimposed to the coarser grid ones, shown in transparency.

Fig.~\ref{fig:2stream_ps_1} is taken at $\omega_{pe}t= 62$, a time at which the differences between $E_{x,P,l0}$ and $E_{x,N,l1}$ are rather pronounced, how it is possible to check by following the red vertical line in Fig.~\ref{fig:Ex2streamANDdensities}, panel a and b. Notice indeed that, comparing the $v_{x}/c$ $vs$ $x/d_{e}$ phase space plots in Fig.~\ref{fig:2stream_ps_1} aI and bI, it is possible to see the cause of the field mismatches at $5 \leq x/d_{e} \leq 6.1$: the inner part of the central electron hole in the overlapping area is moving rightwards with an higher velocity in the coarser grid than in the refined grid, thus producing a shift in the electric field trace. Notice, however, that particles are continuous at the boundaries, as expected.\\
Fig.~\ref{fig:2stream_ps_2}, instead, is taken at a later time, $\omega_{pe}t= 92$, when no mismatch in the electric field is visible between the grids. In this case, it is possible to fully appreciate the advantage that the MLMD simulation offers in terms of resolution over a simulation with the coarse resolution alone. Notice the increased level of details provided by the refined grid phase spaces and how the smallest streams at high, negative $v_{x}/c$ nicely cross the interface, both from the coarser to the refined grid and vice versa, in Fig.~\ref{fig:2stream_ps_2}bI.

\subsection{Testing the reaction of the refined grid to strong driving from the coarser grid: shock simulation}
\label{sec:shock}

\begin{figure}[ht]
\centering
 \includegraphics[width=8cm]{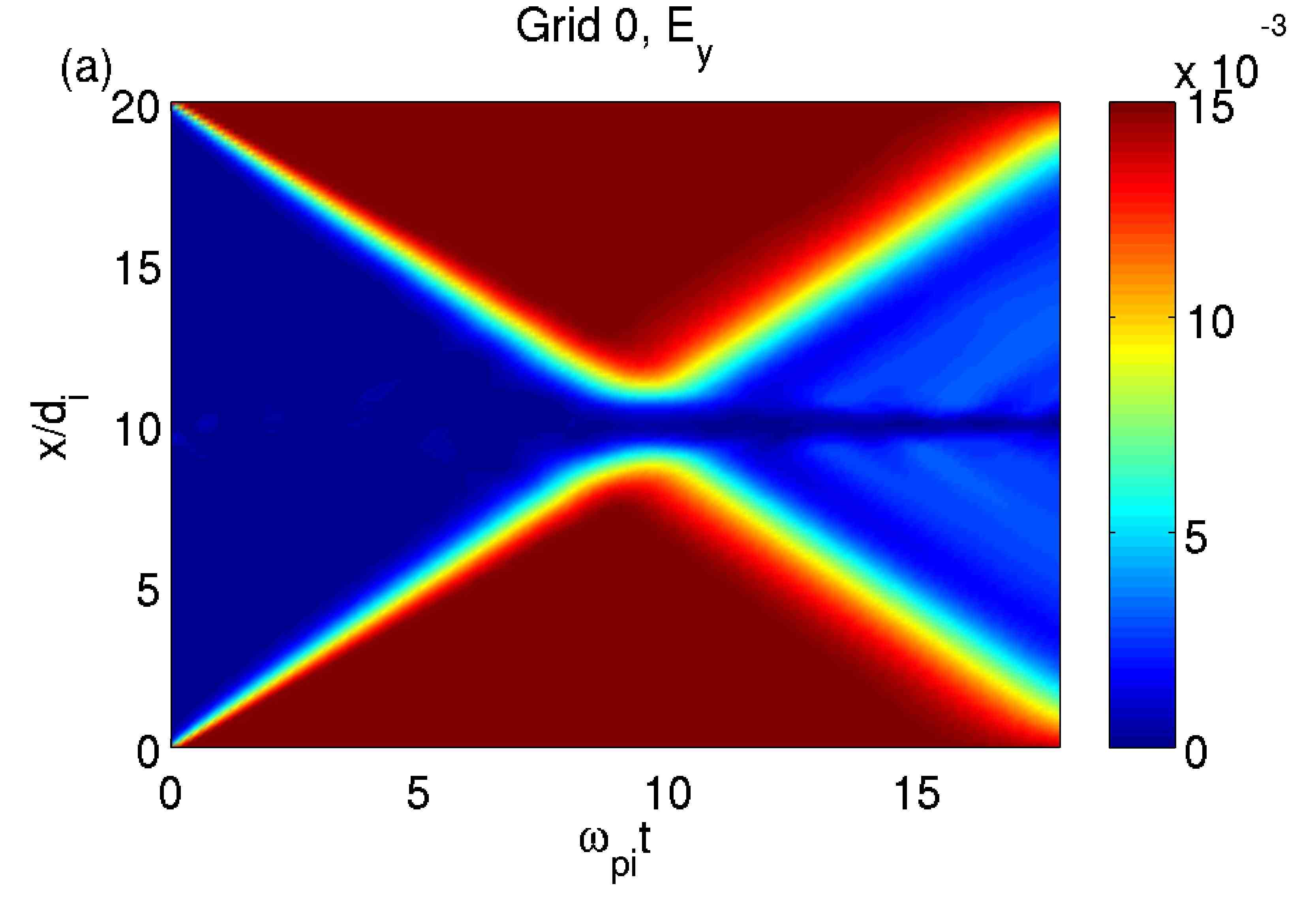}
 \includegraphics[width=8cm]{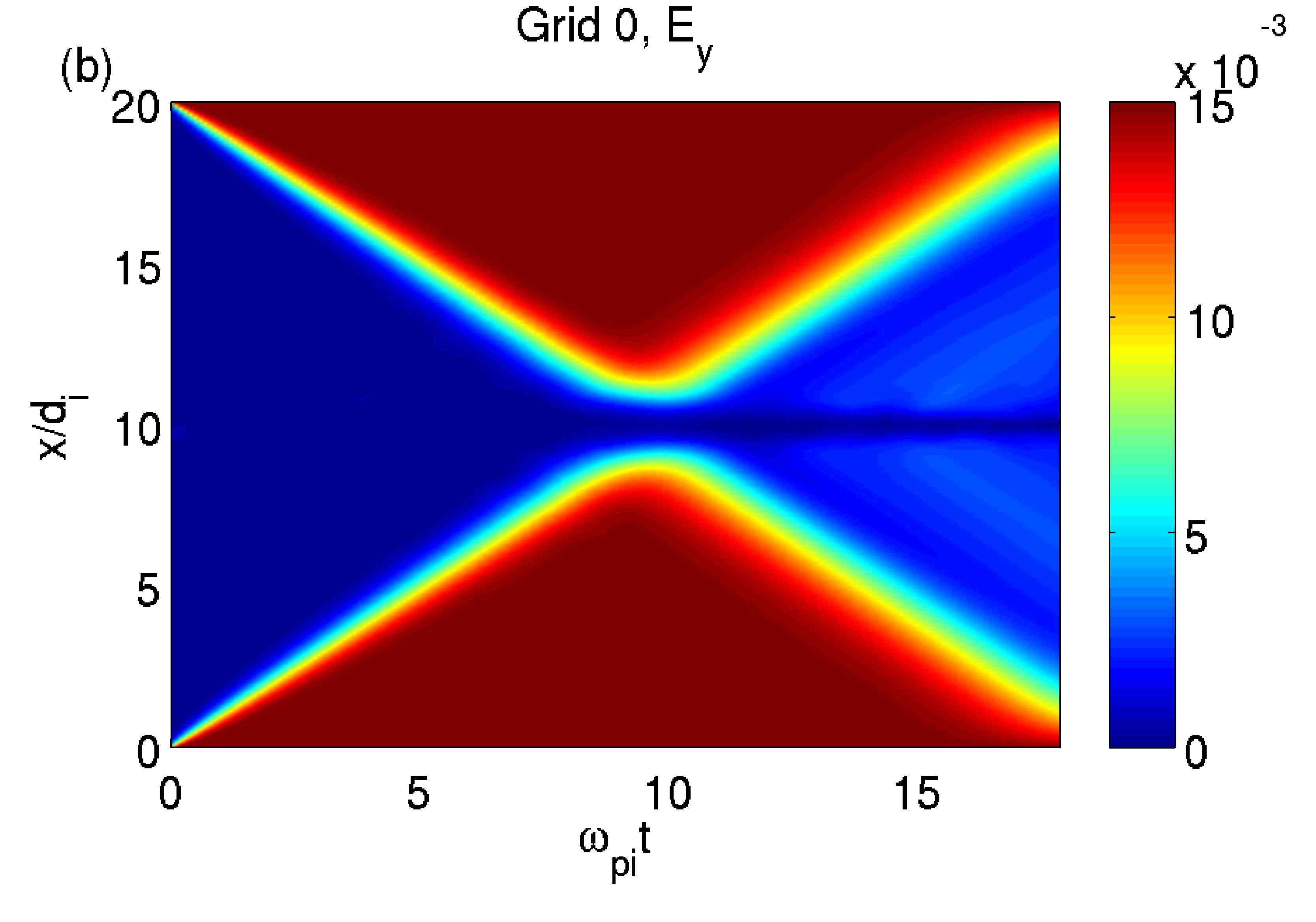}
 \includegraphics[width=8cm]{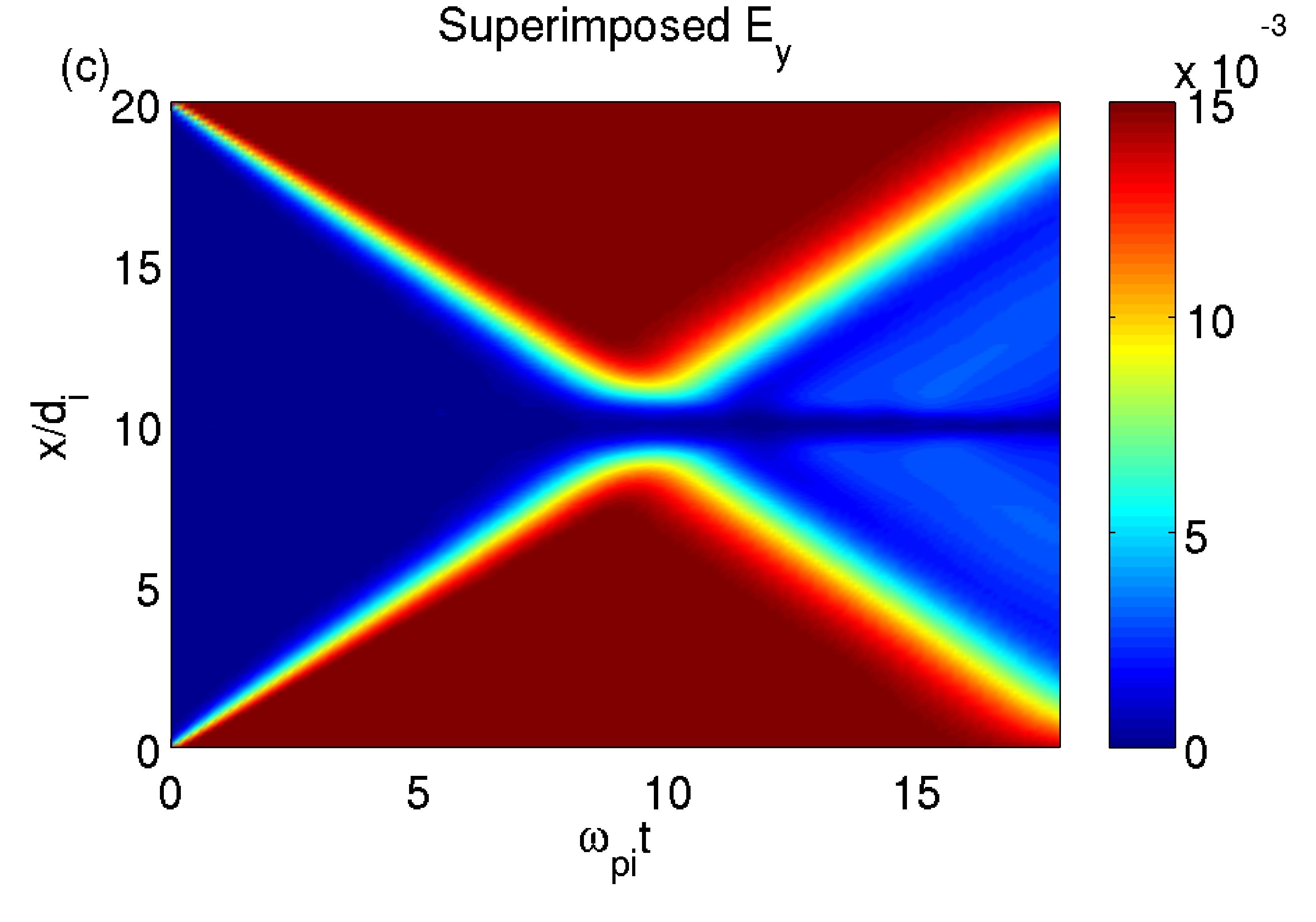}
\caption{Electric field $E_{y}$ evolution when a shock wave is excited at the boundaries of the coarse domain. Panel (a) and (b) depict a reference simulation with the resolution of the coarser and refined grid respectively, panel (c) shows $E_{y,N,g_{l1}}$ superimposed to $E_{y,P,g_{l0}}$ in a Multi Level Multi Domain simulation of shock propagation.}
\label{fig:ShockEy}
\end{figure}

The test problems presented in Sec.~\ref{sec:weibel} and~\ref{sec:2streams} focused on the development of instabilities excited simultaneously on the refined and coarser grids: the issues under investigation were the consistency of the evolution on the two grids and the continuity of particle structures at the grid boundaries.\\
In this case, instead, a shock problem is used to test the effectiveness of the coarser level driving over the refined grid: a shock wave is created at the coarser grid boundaries and then launched towards the center of the domain and thus the refined grid.\\The plasma inside the domain is initially in a Harris state\citep{harris62}, with half width of the current sheet $L_{H}/d_{i}=0.5$ ($d_{i}=c/\omega_{pi}$ is the ion inertial length and $\omega_{pi}$ the ion plasma frequency), a temperature ratio of $T_{i}/T_{e}=5$ and a mass ratio of $m_{i}/m_{e}=25$. The mass and temperature ratio are the same as in the Newton challenge\citep{newton}, while the half width of the current sheet is thinner in order to have a significative particle density only at the center of the coarse domain: the aim is to have the wave excited from the boundary compression (an electric field of arbitrary value $Ey=0.015$ is imposed at the boundaries of the coarser grid) propagating across the domain with light speed. For the same reason, a particle background, which would decelerate the wave propagation velocity to $\bf{v}= \bf{E} \times \bf{B} / B^{2}$, is not included. \\
The MLMD simulation is performed with a coarser grid length of $L_{x, g_{l0}}/d_{i}=20$, $128$ cells for both the coarser and the refined grids, a time step $\omega_{pi}dt=0.1$, a Refinement Factor $RF=4$ and no PRA cells in the refined grid. The refined domain is situated at the center of the coarser grid, at $7.5 \leq x/d_{i} \leq 12.5$.\\
The MLMD evolution is compared with check runs performed with the same grid length and with the coarser and refined level resolution ($128$ and $512$ cells, same grid length) respectively, all the other parameters unchanged. Notice that references runs were not shown (however performed) in the previous cases since the evolution of the systems presented previously is well known and documented.\\
Fig.~\ref{fig:ShockEy} depicts the evolution of $E_{y}$ in the three cases, the coarse (panel a) and fine (panel b) reference cases and the MLMD system (panel c), where the refined grid field $E_{y,N,g_{l1}}$ is superimposed to the coarser grid field  $E_{y,P,g_{l}}$ between $7.5\leq x/d_{i} \leq 12.5$. Notice that in all three cases the wave propagates undisturbed towards the center of the coarse domain and is then reflected. No spurious reflections or front distortion are observable in the MLMD case.

\begin{figure}[ht]
\centering

 \includegraphics[width=8cm]{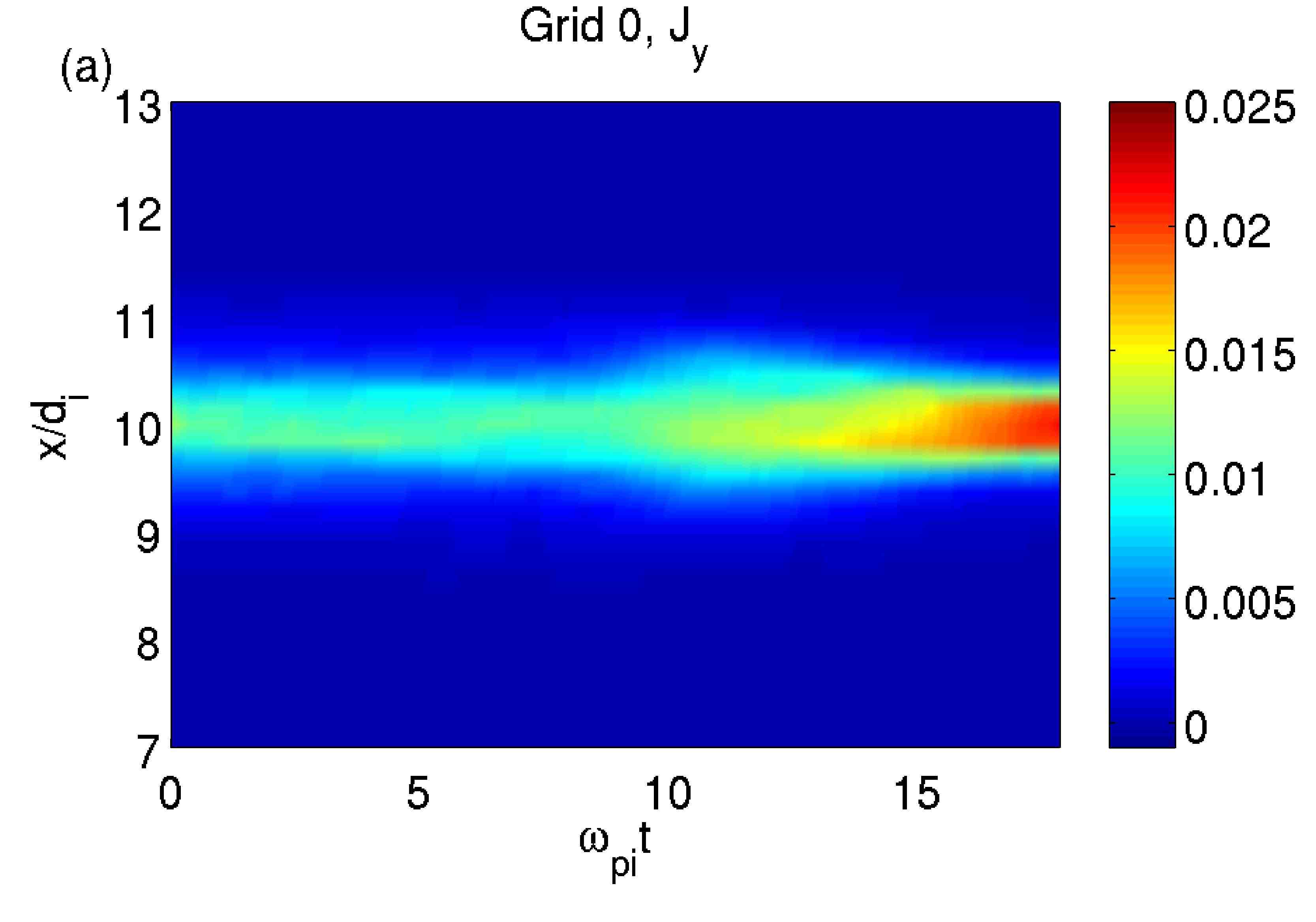}
 \includegraphics[width=8cm]{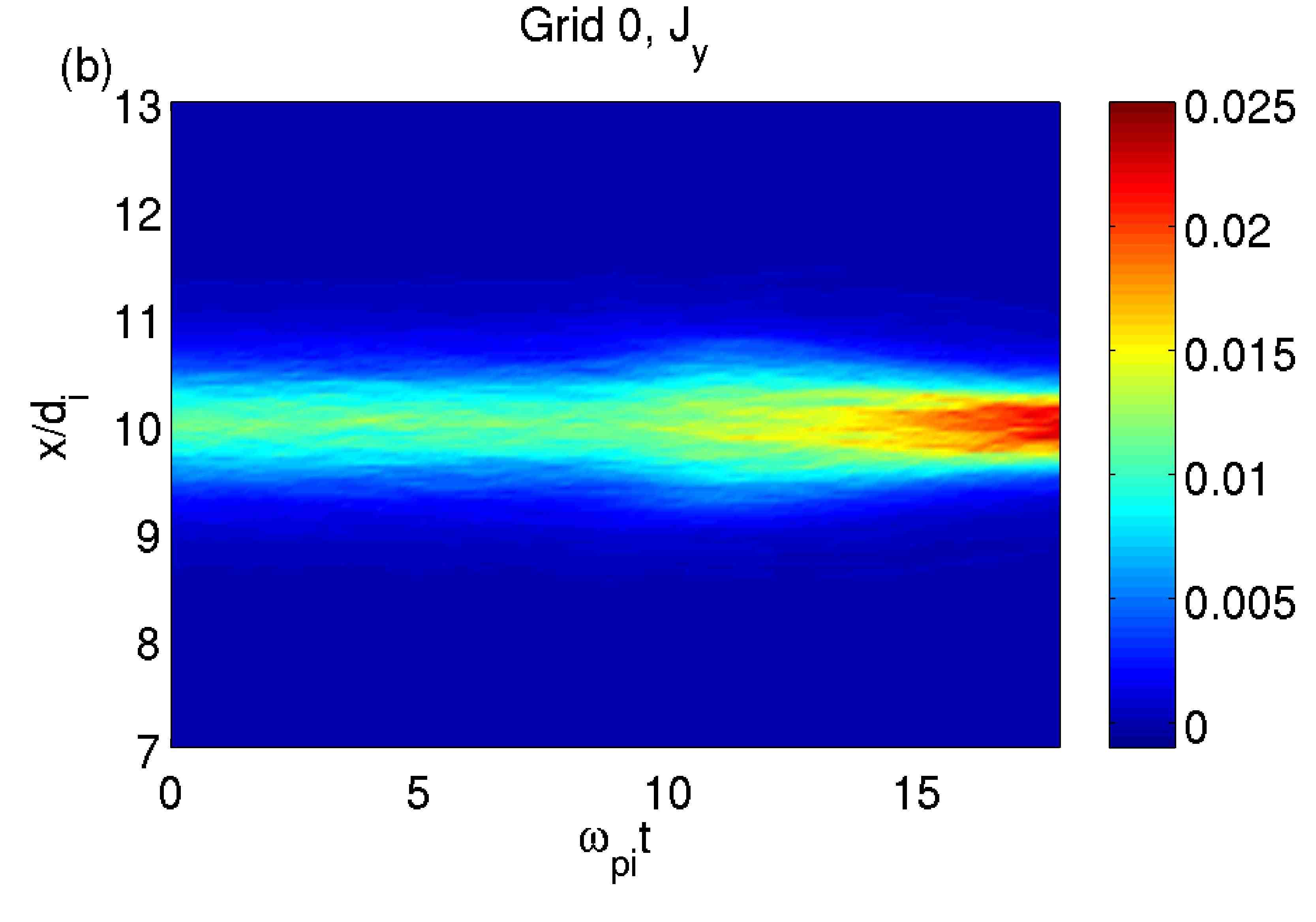}
 \includegraphics[width=8cm]{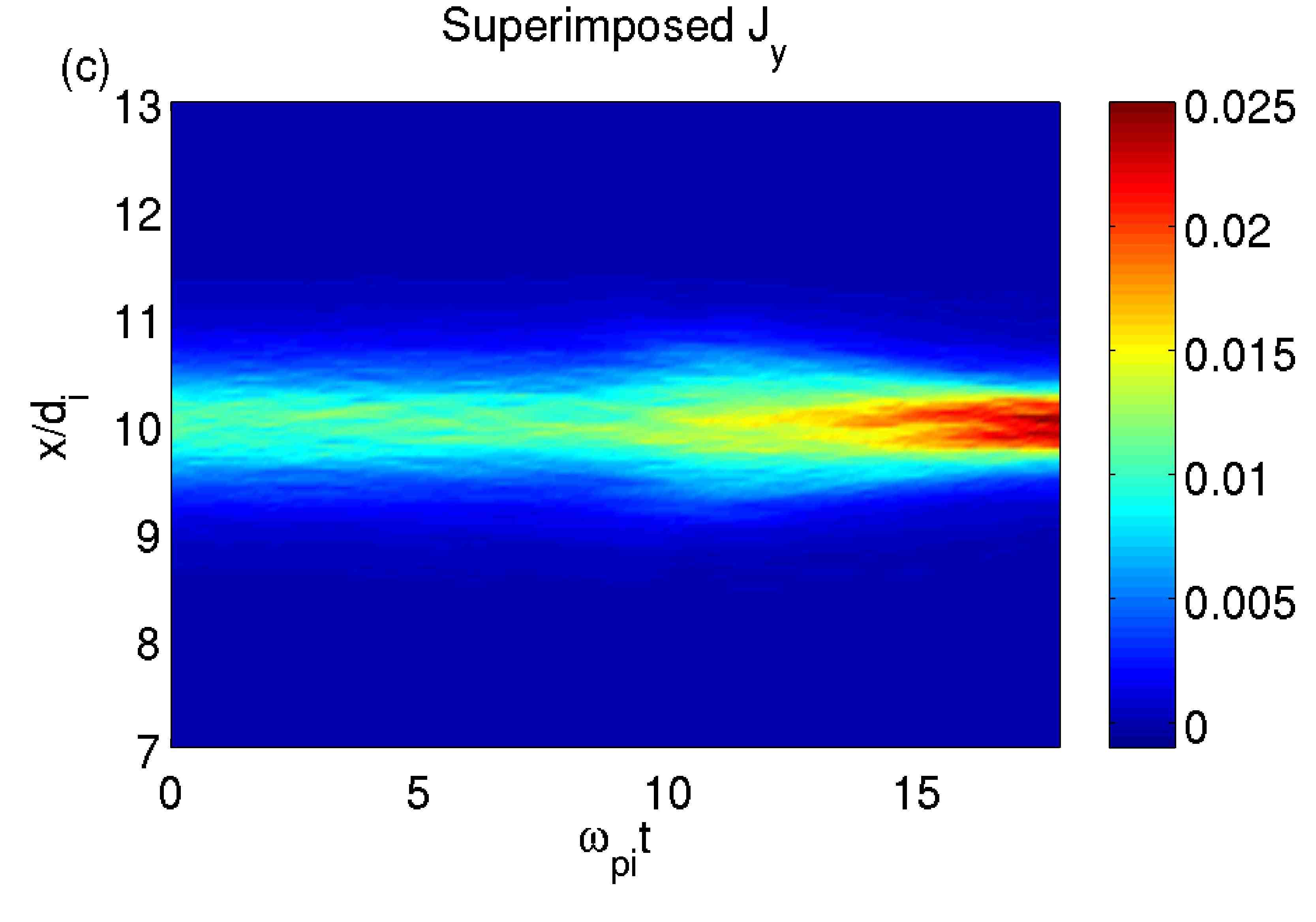}
\caption{Current $J_{y}$ evolution at $7 \leq x/d_{i} \leq 13$ when a shock wave is excited at the boundaries of the coarse domain. Panel (a) and (b) depict a reference simulation with the resolution of the coarser and refined grid respectively, panel (c) shows $J_{y,g_l1}$ superimposed to $J_{y,g_l0}$ in a Multi Level Multi Domain simulation of shock propagation.}
\label{fig:ShockJy}
\end{figure}

Fig.~\ref{fig:ShockJy} shows the $J_{y}$ current evolution again in the three cases, not refined (panel a) and refined (panel b) reference case and MLMD system (panel c). Only the area at $7 \leq x/d_{i} \leq 13$  is shown, since no significative activity is detectable further away from the center of the domain. In all three cases it is possible to observe how the arrival of the shock modifies the current profile starting from $\omega_{pi}t=10$, with an initial thicking of the current profile and a consequent peak. Notice how the MLMD simulation allows to reach the same resolution level for the current as in the refined case in the high density area of the simulation, while removing the need of resolving to an unnecessary level the less relevant areas of the simulation.

\section{Conclusions}
\label{sec:conclusions}

In this paper, a novel solution of the problem of simulating larger systems for longer times and with kinetic scale resolutions is proposed.\\
The novelty of the approach presented is twofold. First, a Multi Level Multi Domain solution is introduced: the simulated levels are fully functional and self-similar and, even when a more refined grid exists for an area, the corresponding coarser area is still simulated completely, with both fields and particles (see Sec.~\ref{sec:particle_shape_solution}). Second, the
framework in which the MLMD technique described here is applied is the Implicit Moment PIC method
(see Sec.~\ref{sec:stability_solution} and references therein for a review of the method).

The choice of adopting a MLMD instead of an AMR approach is one of the possible solutions (see Sec.~\ref{sec:particle_shape}) to the problem of adapting the particle shape size to the grid size. Moreover, having fully functional self-similar levels is a great advantage from the point of view of the ease of programming (with an object-oriented approach, each domain is treated as an instance of the same object) and remarkably simplifies the daunting task of managing the boundaries between the coarser and refined grids.\\
Implicit PIC methods already grant a computational advantage over explicit methods since they notably relax the stability-dictated limits on grid resolutions: they are therefore optimal candidates for AMR and Multi Level evolutions, given the possibility of spanning a notably increased range of time and space steps.\\
The algorithm proposed in the paper prescribes an exchange of information between the coarser and refined grids in three steps (see Sec.~\ref{sec:InterDomainInter}): projection of the refined grid fields from the refined to the coarser grid, interpolation of the refined grid boundaries from the coarser grid fields and particle repopulation at the refined grid boundaries starting from the particle distribution in the coarser grid.\\
The key point in exchanging field information between the grids (i.e., both for projection and interpolation operations) is to consider the grid points in the refined grids as particles in the coarser grid and consequently use for both projection and interpolation the same interpolation functions employed for calculating the momenta from particle positions and velocities.\\
The projection strategies presented have the primary aim to favor the maximum coupling between the levels while preserving fields and particles consistency within the grids. For this reason, the electric fields are projected from the refined to the coarser grids as an average between the native, coarser fields and the refined grid information (Eq.~\ref{eq:proj_E}) and the magnetic fields are recalculated from the projected electric field through the Maxwell-Faraday's equation (Eq.~\ref{eq:proj_B}). Eq.~\ref{eq:proj_E} makes the transition that coarser level particles experience when crossing into the overlap area smoother, while Eq.~\ref{eq:proj_B} is adopted in order to safeguard the validity within the grid of the Maxwell-Faraday equation (Eq.~\ref{eq:maxwell}).\\
The strategy adopted in particle repopulation, then, based on the particle rezoning algorithm of Ref.~\cite{lapentaadapt}, is chosen to favor the grid coupling by exactly reproducing the coarser grid particle distributions at the boundaries of the refined grids.\\
The proposed algorithm is tested in a 1D setting against a series of challenges. In Sec.~\ref{sec:max}, the reaction of the MLMD system to the exchange of information between different levels in absence of plasma activity is tested through the simulation of a Maxwellian plasma. Sec.~\ref{sec:weibel} and~\ref{sec:2streams} focus on the  development of instabilities and the formation of particle structures either fixed in position but grown across the refined-coarser grid boundaries (MLMD simulation of a Weibel instability) or moving across the grid boundaries (MLMD simulation of a two stream instability). Finally, in Sec.~\ref{sec:shock} a shock is excited in the coarser grid and launched across the finer grid to test the effectiveness of the driving at the boundary.\\ 
The results are encouraging: the proposed MLMD algorithm proves to grant an excellent coupling between refined and coarser levels (Sec.~\ref{sec:max}) and not to impair the development of instabilities and particle structures across the grids (Sec.~\ref{sec:weibel} and~\ref{sec:2streams}). Moreover, the shock test shows that even an extreme case, like a wave propagating at light speed across the coarser-refined grid boundaries, is correctly handled by the system and, even more relevantly, that the aim of the MLMD system is fulfilled. In cases when only a small section of the simulation requires enhanced resolution, while the rest may allow the use of a coarser grid, the algorithm proposed proves capable of correctly driving the refined grid towards the evolution expected in a simulation fully performed with the refined level resolutions, without wasting computational resources in the areas which requires less resolution. \\
Future work will focus on the application of the proposed algorithm to the operative implicit PIC code Parsek2D~\citep{ipic} and on the quantitative assessments of the performance increase granted by the MLMD method when used in a parallel implementation.

\section*{Acknowledgments}
The present work is supported by the Exascience Intel Lab Europe, by the
Onderzoekfonds KU Leuven (Research Fund KU Leuven) and by
the European Commission's Seventh Framework Programme
(FP7/2007-2013) under the grant agreement no. 263340 (SWIFF project, www.swiff.eu) and No. 218816 (SOTERIA project, www.soteria-space.eu)


\end{document}